\documentclass[twocolumn]{aastex631}

\usepackage{multirow}
\usepackage{hyperref}

\newcommand{\Ha}{H$\,\!\alpha$}
\newcommand{\Db}{D4000}
\newcommand{\Msun}{$M_\odot$}

\newcommand{\kmsM}{\,km\,s$^{-1}$\,Mpc$^{-1}$}
\newcommand{\Fring}{F_\mathrm{ring}^\mathrm{WH\alpha}}

\newcommand{\citeaposauthor}[1]{\citeauthor{#1}'s}

\defcitealias{Tous+2020}{TSP20}

\submitjournal{ApJ; accepted for publication; v. 2.0}

\shorttitle{Star-forming rings in S0 galaxies}
\shortauthors{Tous et al.}

\setwatermarkfontsize{50pt}

\graphicspath{{./}{figures/}}

\begin{document}

\title{The origin of star-forming rings in S0 galaxies}

\correspondingauthor{J.\ L.\ Tous}
\email{jtous@fqa.ub.edu}

\author[0000-0002-1569-0468]{J.\ L.\ Tous}
\affiliation{Departament de F\'\i sica Qu\`antica i Astrof\'\i sica i Institut de Ci\`encies del Cosmos (ICCUB), Universitat de Barcelona. \\ Mart\'{\i} i Franqu\`es 1, E-08028 Barcelona, Spain}

\author{H.\ Dom\'\i nguez-S\'anchez}
\affiliation{Instituto de Ciencias del Espacio (ICE-CSIC), Campus UAB. Can Magrans s/n, E-08193 Barcelona, Spain}
\affiliation{Centro de Estudios de Física del Cosmos de Arag\'on (CEFCA). Plaza San Juan, 1, E-44001, Teruel, Spain}

\author[0000-0002-1633-0499]{J.\ M.\ Solanes}
\affiliation{Departament de F\'\i sica Qu\`antica i Astrof\'\i sica i Institut de Ci\`encies del Cosmos (ICCUB), Universitat de Barcelona. \\ Mart\'{\i} i Franqu\`es 1, E-08028 Barcelona, Spain}

\author[0000-0002-5640-9791]{J.\ D.\ Perea}
\affiliation{Instituto de Astrof\'\i sica de Andaluc\'ia (IAA-CSIC). Glorieta de la Astronom\'\i a s/n, E-18008 Granada, Spain}

\begin{abstract}
Spatially resolved integral field spectroscopic maps in a sample of $532$ S0 galaxies from the MaNGA survey have unveiled the existence of inner rings ($\langle R\rangle\sim 1\,R_\mathrm{e}$) betraying ongoing star formation in a number of these objects. Activity gradients averaged over bins of galactocentric radius up to $\sim 1.5\,R_\mathrm{e}$ have been measured in the subspace defined by the first two principal components of the optical spectra of these galaxies. We find that the sign of the gradients is closely related to the presence of such rings in the spectral maps, which are especially conspicuous in the equivalent width of the \Ha\ emission line, EW(\Ha), with a fractional abundance ---$21$--$34\%$--- notably larger than that inferred from optical images. While the numbers of S0s with positive, negative, and flat activity gradients are comparable, star-forming rings are largely found in objects for which quenching proceeds from the inside out, in good agreement with predictions from cosmological simulations studying S0 buildup. Assessment of these ringed structures indicates that their frequency increases with the mass of their hosts, that they have shorter lifetimes in galaxies with ongoing star formation, that they may feed on gas from the disks, and that the local environment does not play a relevant role in their formation. We conclude that the presence of inner rings in the EW(\Ha) is a common phenomenon in fully formed S0s, possibly associated with annular disk resonances driven by weakly disruptive mergers preferentially involving a relatively massive primary galaxy and a tiny satellite strongly bound to the former.
\end{abstract}


\section{Introduction}
\label{S:intro}

Galaxies of Hubble type S0 (also known as lenticulars) are a heterogeneous population with a wide range of physical properties. They are also the only morphological class that is abundant in both low- and high-density environments, suggesting that different physical mechanisms participate in their formation \citep[e.g.][]{WE12}. In recent years, it has become clear, through observations and both semianalytical and numerical modeling, that minor mergers of a disk galaxy with a gas-rich satellite or even major mergers between star-forming (SF) spirals taking place in the field or in small galaxy groups can result in lenticular remnants with a strong structural bulge--disk coupling \citep{Ber+11,Pri+13,Que+15,Tap+17,EliM+18,Mas+20}. Conversely, there is also the more conventional formation pathway related to dense environments that involves the removal of the intragalactic gas through hydrodynamic processes that, with the help of other mechanisms such as harassment or strangulation \citep{LTC80,Nul82,Moo+99}, is capable of transforming spiral disks infalling into large galaxy aggregations ---usually as part of a group--- into stellar systems with a lenticular structure \citep{SSS92,Dre+97,Fas+00,QMB00,CK08,Pog+09,DOMB15}.

The defining characteristic of S0 galaxies within the disk-galaxy family is the absence of visible spiral arms. However, these objects can show other kinds of dynamical instabilities, such as bars and rings. But just as S0 is the disk-galaxy class where the former features are less abundant and weaker \citep{AMAC09,WE12}, the same is not true for galaxy-scale rings, whose fractional abundance is larger among early-type disks, as shown for instance in the statistical study of stellar rings in present-day galaxies by \citet{Com+14} based on near-infrared data \citep[see also][]{Fer+21}. This negative trend between the frequencies of the two types of transient structural components linked to host morphology, as well as the finding by these and other authors that rings are only partially favored in barred galaxies and not necessarily aligned with bars, suggests that bars and rings may not always be intimately connected, leading to the idea that perhaps not all in-plane rings are disk instability phenomena that originated in bar resonances \citep[][]{Ath+82,BC96}.

Another relatively novel aspect unveiled by recent studies of S0 galaxies is the finding that they show a duality in their physical properties \citep{Xiao+2016,mckelvie+2018,HDS+Bernardi+2020}. This has been put on a firm ground by the statistical analysis of the integrated optical spectra of a sample of more than $68,\!000$ present-day ($z\lesssim 0.1$) lenticulars carried out by \citeauthor{Tous+2020}~(\citeyear{Tous+2020}; hereafter \citetalias{Tous+2020}) using data from the Sloan Digital Sky Survey \citep[SDSS-III]{Ala+15}. The principal component analysis (PCA) of the SDSS spectra demonstrated that in the PC1--PC2 plane of the first two principal components (PCs) S0 galaxies can be grouped into two main activity classes, named the passive sequence (PS) and the active cloud (AC), separated by a slim dividing zone that constitutes the transition region (TR). Thus, although the majority of the local S0 population belongs to the PS, which conforms to the traditional conception of these objects as quiescent galaxies, approximately the quarter of those found in the AC may show star formation rates (SFRs) and spectral characteristics comparable to those measured in late spirals. This diversity of properties exhibited by S0 galaxies is an expected result in objects that follow the aforementioned multiplicity of formation channels \citep[see also][]{Welch+2003,Morganti+2006,Crocker+2011,Barway+2013,Deeley+2020}. Interestingly, high-resolution far-UV imaging of early-type galaxies from the Hubble Space Telescope/Advanced Camera for Surveys shows that in S0 galaxies with strong UV excess the UV emission is frequently organized in concentric ring- or arm-like structures \citep{SR10,Sal+12}.

In this work, we apply the same PCA technique of \citetalias{Tous+2020} to study the spatially resolved optical spectra of a sample of $532$ galaxies representative of the local S0 population drawn from the integral field spectroscopic (IFS) survey Mapping Nearby Galaxies at Apache Point Observatory \citetext{MaNGA, \citealt{Bundy+2015}; \S~\ref{S:data}}, with a focus on aspects related to their star formation activity. In  \S~\ref{S:gradients} we provide a way to typify the radial spectral profiles of these objects that in \S~\ref{S:dim_reduction} are synthesized in the PC1--PC2 subspace, which facilitates their subsequent analysis. The intercomparison of the systematized spectral profiles and several spatially resolved spectral indicators sensitive to the distribution of the dense interstellar gas and the mean age of the stellar component reveals in \S~\ref{S:rings} that the typology of the radial gradients of activity in S0 galaxies is closely linked to the presence of SF rings. Aided by this finding and the behavior shown by the fraction of S0 galaxies hosting rings in the equivalent width of \Ha\ emission, EW(\Ha), with the global properties of stellar mass and environmental density (\S~\ref{S:origin_of_rings}), we postulate in \S~\ref{S:scenario} that these SF annuli are transient structures mostly driven by scarcely disruptive mergers with tiny mass ratios that could well be significantly less than 1:10. This and other important outcomes of this work are summarized in \S~\ref{S:conclusions}. The manuscript also includes two appendices. In Appendix~\ref{A:PCA}, we provide a short outline of the PCA methodology and the resulting spectral classes, while in  Appendix~\ref{A:all_maps}, we present the montages of optical images and spectral maps for S0 galaxies with linear PC1--PC2 radial profiles that are published in the online version of the paper, emphasizing those objects harboring an unequivocal ring in the EW(\Ha) map. We note that the relatively large number of systems studied, combined with the high rate of detection of SF rings, makes this S0 sample from MaNGA the largest collection of ringed galaxies ever identified in the local universe by spectroscopic observations.

All cosmology-dependent quantities used in the paper assume a flat concordant $\mathrm{\Lambda}$CDM universe with parameters $H_0 = 70$ \kmsM and $\mathrm{\Omega_m} = 1 - \mathrm{\Omega_\Lambda} = 0.3$.

\section{Sample selection}
\label{S:data}
We use data from the Fifteenth Data Release of the Sloan Digital Sky Surveys \citep{Agu+19}, which includes the first 4621 MaNGA data cubes on unique galaxies, from the $\sim 10,\!000$ initially planned. This survey provides spatially resolved spectroscopic information on a statistically representative sample of nearby ($\tilde{z}\sim 0.03$) galaxies out to about 1.5 effective radii in the $r$ band, $R_\mathrm{e}$, over the bandpass $3600$--$10000$ \AA\ with a spectral resolving power $R=\lambda/\mbox{FWHM}\sim 2000$ \citep{Sme+13} and a spatial resolution of $\sim 1\,$kpc at the median redshift $\tilde{z}$ of the observations. With the only exception of the total stellar mass $M_\ast$, for which we use the estimates listed in the GALEX--SDSS--WISE Legacy Catalog 2 \citep[][]{SBL18} derived from UV-to-mid-IR spectral energy distribution fitting, we rely on the physical parameters from the continuum and ionized emission lines provided by the Data Analysis Pipeline (DAP) of MaNGA \citep{Westfall+2019}. With them, we build for the target galaxies resolved maps of the flux intensity and EW of their \Ha\ line emission, which in SF galaxies inform, respectively, on the spatial distribution of the ionized gas tracing instantaneous star formation activity and on the specific star formation rate (SSFR), as well as maps of the break at 4000 \AA, \Db, which is a proxy for the age of the stellar component and therefore also probes star formation but on longer timescales. Details of the emission-line measurements by the DAP can be found in \cite{Westfall+2019}. In short, the spaxels are Voronoi binned to reach a signal-to-noise ratio (S/N) of at least $10$, and kinematic measurements, including the line-of-sight velocity\footnote{The DAP uses a stellar-template library constructed by hierarchically clustering the MILES stellar library \citep{Fal11} into a set of 42 composite spectra, termed the MILESHC library, to measure the stellar kinematics; only the first two velocity moments ($v$ and $\sigma_v$) are provided.}, are computed. In this work we use the HYB10 maps, for which emission-line and spectral-index measurements are performed on the individual spaxels, after subtracting the best-fitting stellar continuum. Likewise, we have retrieved the radius, derived from the semimajor-axis elliptical polar coordinates, in units of the effective radius of the galaxies reported by the NASA-Sloan Atlas catalog \citep{Blanton+2011}, $R^\prime = R/R_{\rm e}$. Throughout this work, we use MARVIN \citep{Che+19} as a tool to access and manipulate all MaNGA data.

Lenticulars are identified using the deep learning-based morphological classification of \citet{Fischer+2019}, founded on the methodology detailed in \citet{HDS+2018}, who provide this information for MaNGA galaxies in terms of two parameters: a numerical Hubble stage, $T$, and the probability, $P_{\mathrm{S0}}$, that an early-type-looking object is truly an S0. After imposing $T \leq 0$ and $P_{\rm S0} > 0.7$, we are left with a sample of $648$ galaxies that reduce to $532$ bona fide S0s, once we remove the problematic cases flagged by the MaNGA data reduction pipeline, as well as those galaxies for which we have detected, after a stringent visual inspection of the SDSS images, a hint of spiral structure, and/or contaminating foreground stars, companion galaxies, or both within the MaNGA's field of view.

The PCA-based classification (see Appendix~\ref{A:PCA} for details of this methodology) of the selected galaxies has been inferred by cross-matching them with the single-fiber spectral sample of S0 defined in \citetalias{Tous+2020}, while the few objects without a counterpart in this dataset have been classified from scratch using the first two PCs of their single-fiber SDSS spectra. This exercise shows that our MaNGA S0 sample has $343$ galaxies that are PS members and $150$ that belong to the AC. The number of TR objects (39) is not large enough to perform a statistically significant analysis separately for this group, so from now on we will include them under the AC designation, since the behavior of their EW(\Ha) profiles is more similar to that shown by the members of this activity class than that of PS-type objects (see \S~\ref{S:gradients}).

\section{Data cube processing}
\label{S:dim_reduction}

To avoid the complications inherent in handling the relatively large number of spaxels in each data cube and the disparity in their areal coverages ---the MaNGA integral field units (IFUs) are bundles of between 19 and 127 fibers to optimize the match with the apparent sizes of target galaxies--- we choose to stack those with S/N $>5$ into circularized radial bins in the dimensionless variable $R^\prime$ within the interval $[0,1.5]$. Before we proceed with the stacking, the spectra are corrected for Galactic extinction and cleaned from masked regions, glitches, and sky lines, discarding those in which these regions represent more than $10\%$ of the flux array. We also remove the recessional velocity of each galaxy and the mean rotation speed from each spaxel. Finally, we normalize the spectra using Equation [3] from \citetalias{Tous+2020} and weight them by the S/N of the continuum to generate the mean spectrum of each radial bin, provided that more than $20\%$ of the spaxels in that bin have a spectrum that fulfills the aforementioned quality conditions.

\begin{figure*}[ht!]
\centering
  \gridline{\fig{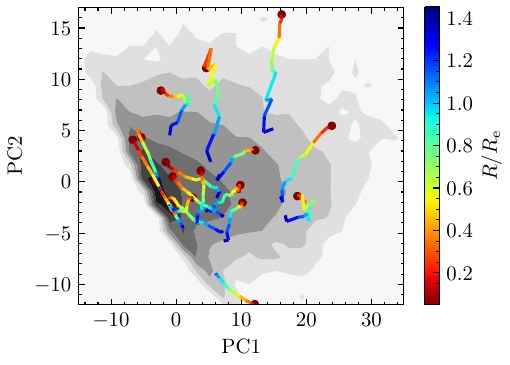}{0.49\textwidth}{} \fig{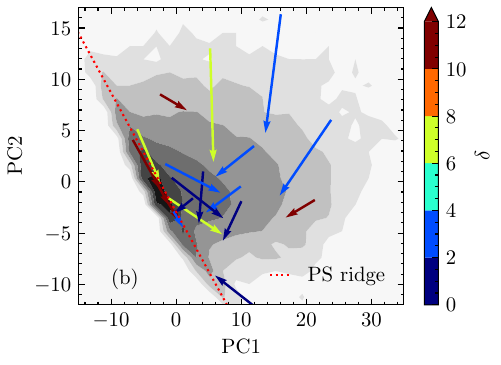}{0.48\textwidth}{}}
\vspace*{-1.5\baselineskip}
\caption{Left: radial profiles of a random subset of S0 galaxies in the subspace defined by the first two PCs of the entire sample. The composite spectra in each radial bin are colored according to the galactocentric distance scaled by the effective radius of the galaxies, with brown dots indicating the position of the central (innermost) bins. Right: vectorization of the spectral profiles. The colors show the value of $\delta$, one of the two parameters used to evaluate its feasibility (see text). The brown arrows identify profiles with $\delta\geq 10$, unsuitable for vectorization. The red dotted straight line shows the location of the PS ridge. In both panels, the background grayscale contours represent equally spaced logarithmic densities drawn from the sample of $\sim 68,\!000$ nearby ($z\lesssim 0.1$) lenticulars explored in \citetalias{Tous+2020}.}
\label{fig:sample_profiles}
\end{figure*}

Next, we reduce the multiple dimensions of the flux arrays that construct the composite spectra in each radial bin to their projections into the first two PCs inferred from the global spectra. The left-hand panel of Figure~\ref{fig:sample_profiles} shows the spectral radial profiles obtained by applying this dimensionality reduction to a random subset of our sample of 532 S0 galaxies.

The fact that a great deal of the radial profiles in the PC1--PC2 subspace show an approximate linear behavior allows us to go one step further to simplify the analysis of the MaNGA data without missing essential information. Thus, we have best-fitted all these spectral profiles by means of straight lines that have then been converted into vectors using the projections of the points corresponding to the innermost and outermost radial bins on such lines to set, respectively, the tail and tip of the arrows. However, as shown in  Figure~\ref{fig:sample_profiles}, there are some vectors, colored in brown in the right-hand panel, that are a poor representation of the spectral radial profiles they intend to replace. To objectively measure the goodness of this approximation and therefore to identify those spectral profiles that can be adequately vectorized, we introduce two empirical parameters: the disorder, $\delta$, that determines the extent to which the projections of the PC1 and PC2 coordinates of the radial bins into the model lines follow the right sequence of radial distances; and the entanglement, $\epsilon$, which is a measure of the lack of linearity shown by the radial profiles in the PC1--PC2 plane inferred from the ratio between the second and the first singular values of the PC1 and PC2 coordinates of the spectral profiles. After some trial and error, we came up with $\delta \geq 20$ or ($\delta \geq 10$ and $\epsilon \geq 0.05$) as the range(s) of these parameters for which vectorization should be avoided. Applying these thresholds, we exclude about $16\%$ of the S0s in our sample, $62$ from the PS region and $24$ from the AC(+TR) one, which represent $\sim 18\%$ and $13\%$ of the corresponding subsets. The details of the whole procedure of dimensionality reduction of the MaNGA data cubes and vectorization of the spectral radial profiles will be presented in Tous et al., in preparation.

\begin{figure*}[ht!]
\centering
  \gridline{\fig{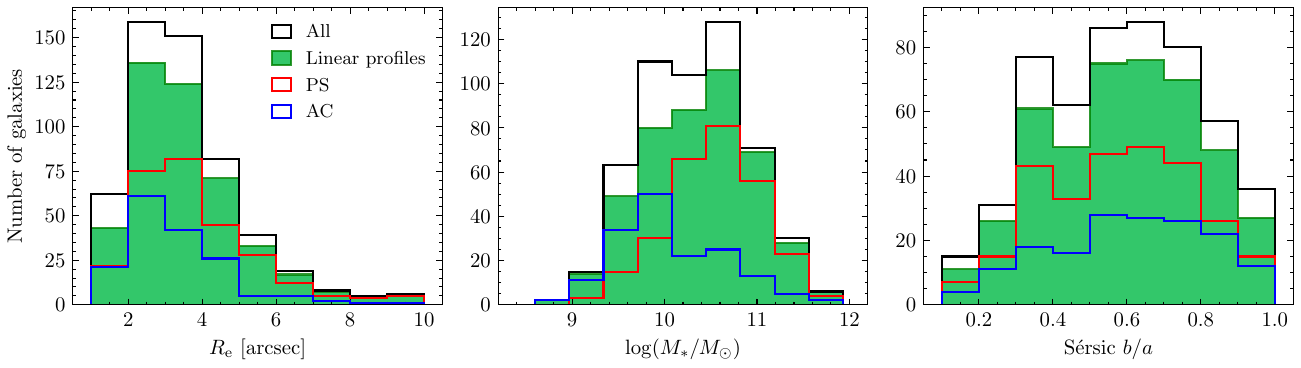}{\textwidth}{}}
\vspace*{-1.5\baselineskip}
\caption{Histograms comparing the effective radius (left), stellar mass (middle) and observed axis ratio (right) of the S0 sample before (empty black histograms) and after (green filled histograms) removing the galaxies with nonlinear PC1--PC2 radial profiles. The red and blue empty histograms show the distributions of these properties for, respectively, the objects of the PS and AC spectral classes with linear profiles.}
\label{fig:sample_properties}
\end{figure*}

As can be deducted from Figure~\ref{fig:sample_properties}, the removal of the galaxies with nonlinear PC1--PC2 radial profiles from the original dataset (empty black histograms) does not introduce meaningful differences in the distributions of effective radius and stellar mass of the reduced sample (green filled histograms). Likewise, this figure shows that, despite spanning the same ranges in these two physical properties, S0 galaxies that belong to the AC class (empty blue histograms) are, on average, slightly smaller and less massive than their PS counterparts (empty red histograms). As expected, no significant biases are observed in the distributions of apparent inclinations. Furthermore, all the spectral images of the S0s with linear spectral profiles extend beyond the FWHM of the point-spread function (PSF), with a median $\sigma_{\rm PSF}/R_{\rm max}$ of $0.23$, and have outermost radial bins that extend to $R^\prime_{\rm max}\sim 1.5$, except for four objects that have $R^\prime_{\rm max}\lesssim 1$.

\section{Activity/Star formation gradients}
\label{S:gradients}

To facilitate the analysis of the inner spectral profiles of S0 galaxies, we introduce a new dimension in their classification that accounts for their orientation relative to the PS, the narrow and well-defined band that demarcates the locus of fully quiescent galaxies in the PC1--PC2 plane. We know from the work by \citetalias{Tous+2020} \citep[see also][]{JimP+22} that the distance to the main ridge of the PS measured in this subspace shows a strong positive correlation with the EW(\Ha). Since this last parameter is an observational proxy for the SSFR, the behavior of the spectral profiles with respect to the PS will inform about the sign of the radial gradient of star formation in the target galaxies. We want to see if it bears any relationship with the classes of activity diagnosed by PCA.

This new dimension is the polar angle, $\theta$, of the vectorized spectral profiles measured counterclockwise from the PS ridge (red dotted line in the right-hand panel of Fig.~\ref{fig:sample_profiles}). We define three orientation tiers. Objects with $5^\circ < \theta < 175^\circ$ are those in which the distance to the PS ridge, and hence the strength of the EW of the \Ha\ line, increase with galactocentric distance, i.e., have positive radial gradients up to $\sim 1.5\,R_\mathrm{e}$. This can be understood as an indication that the star formation activity in these galaxies, irrespective of its global level, is more suppressed in the central regions. Such a group of inside-out (IO) quenched S0s has $113$ members ($25\%$ of the objects with vectorizable spectral profiles), 71 of these being PS systems ($26\%$ of the galaxies of this activity class) and 42 AC ($25\%$ of them). There are $116$ S0s ($26\%$) with $185^\circ < \theta < 355^\circ$ for which the radial gradient of EW(\Ha) is negative. It is remarkable that only nine of the galaxies whose star formation is being suppressed from the outside in (OI) are categorized as PS, while $107$ are AC members ($3\%$ and $65\%$ of their respective activity classes\footnote{The fact that about $40\%$ of the few systems tagged TR show OI-quenched spectral profiles corroborates our decision to include them within the AC subset.}). The remaining 217 galaxies (201 PS and 16 AC) have $\theta\in[-5^\circ,5^\circ]\cup[175^\circ,185^\circ]$ and therefore spectral profiles that run basically parallel to the PS ridge; or, in other words, they have essentially uniform SSFRs. The vast majority correspond to fully retired objects.

All IO-quenched galaxies show a central depression in the ionization level relative to outer regions of the disk. However, not all members of this class are necessarily centrally quiescent, since for objects of type AC, the EW(\Ha) of the spaxels located at small galactocentric radii remains, with few exceptions, above 6 \AA. For their part, the S0--PS at $R'\lesssim 0.5$ tend to be dominated by low-ionization emission-line regions with values of the EW(\Ha) below $3\,$\AA\ that go above this threshold at larger radii. Some of these latter systems may be associated with galaxies with central low-ionization emission-line regions (cLIERs) included in the spatially resolved spectral classification by \citet{Bel+17} and with galaxies in a quenching stage of central quiescence identified in the Calar Alto Legacy Integral Field Spectroscopy Area (CALIFA) survey by \citet{Kal+21}, while others seem to fit better into the \citeaposauthor{Kal+21} category of `nearly retired galaxies', as they show global averages of the EW(\Ha) equal to or less than 3 \AA. On the other hand, our S0--PS with flat spectral profiles resemble the extended LIER (eLIER) and fully retired galaxy classes identified in both works. Surprisingly, we do not find obvious counterparts in these classifications for our S0--AC with OI quenching.

As to the typical value of the average EW(\Ha) per galaxy for each one of the six different spectral categories of S0s that result from combining the activity (PS/AC) and quenching (IO/OI/flat) classes, we have, in descending order: $\langle\mbox{EW(\Ha)}\rangle = 31.6$~\AA\ for OI+AC lenticulars, $20.0$~\AA\ for IO+AC, $8.7$~\AA\ for flat+AC, $4.8$~\AA\ for IO+PS, $1.8$~\AA\ for OI+PS, and $1.6$~\AA\ for flat+PS. This shows that all S0--ACs, regardless of their quenching type, can be considered as SF S0s, since their representative values of $\langle\mbox{EW(\Ha)}\rangle$ are well above the minimum threshold of 6~\AA\ expected when gas excitation is powered by \ion{H$\!\!$}{2} regions \citep[e.g.][]{Cid+11}. It can also be seen that, within this activity class, the highest SSFRs occur on objects with OI-quenched spectral profiles, which are also the most numerous. This indicates, in good agreement with the findings by \citet{Rat+22}, that the characteristic image of S0--ACs, when considered as a whole, would be that of early-type disks with centrally dominated star formation that reaches values of the SSFRs perfectly comparable to those observed in SF spirals, but that is less extended in galactocentric radius. Furthermore, the general quiescence of the PS lenticulars is confirmed, especially for objects with OI and flat spectral profiles, for which the condition $\langle\mbox{EW(\Ha)}\rangle < 3$~\AA\ indicates that the source of the ionized gas is likely hot, low-mass, evolved stars, whereas the intermediate emission-line values of the IO-quenched S0--PS could reflect a mixture of ionization sources \citep{Lac+18}. 

\begin{figure*}[ht!]
\gridline{\fig{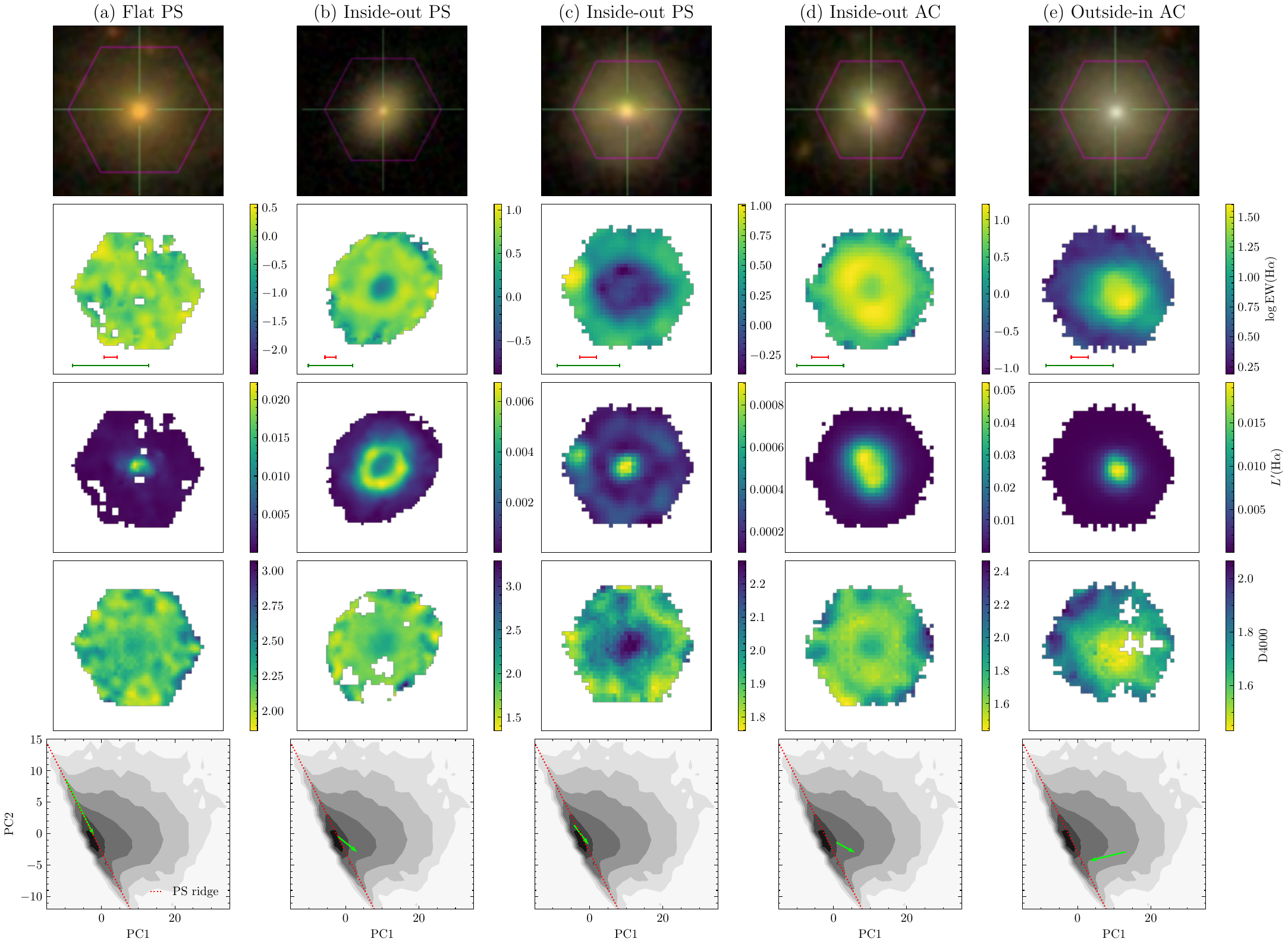}{0.97\textwidth}{}}
\vspace*{-1.5\baselineskip}
\caption{Examples of spectral maps and spectral profiles of S0 galaxies. The second, third, and fourth rows show, in this order, maps of the EW(\Ha), $L^\prime$(\Ha), and \Db\ spectral indices. Only spaxels with all quality flags equal to zero are drawn. The green and red horizontal lines included in the EW(\Ha) panels indicate, respectively, the galaxies' Petrosian half-light diameter in the $r$ band and $2\,\sigma_{\rm PSF}$. The bottom row shows the vectorized spectral profiles of the galaxies in the PC1--PC2 subspace. The true color images of the galaxies are shown in the first row, with the overlapping violet hexagonal frames depicting the footprint of MaNGA's bundles (images' source: \href{https://data.sdss.org/sas/dr15/manga/}{https://data.sdss.org/sas/dr15/manga/}). Galaxies in columns (b), (c), and (d) show clear rings in EW(\Ha) (i.e., they have ring scores $r\geq 6$), whereas no rings are observed for the galaxies in columns (a) and (e) ($r < 3$).}
\label{fig:rings}
\end{figure*}

\section{Patterns in the spectral maps}
\label{S:rings}

In this section, we investigate the dominant patterns described by the spatially resolved stellar and ionized gas distributions in S0 galaxies and their possible links with the spectral categories that we have defined in the previous sections. We use for this endeavor maps of the logarithm of the EW of the \Ha\ line measured in angstrom and of the flux intensity of this same line in units of $10^{-17}$\,erg\,cm$^{-2}$\,s$^{-1}$\,\AA$^{-1}$\,spaxel$^{-1}$ times the heliocentric redshift squared of the galaxy, so it provides an estimate of the \Ha\ luminosity, $L^\prime$(\Ha)\,$=kL$(\Ha), except for the constant of proportionality $k=(H_0/c)^2/4\pi $, as well as of the dimensionless \Db\ index.

\addtolength{\tabcolsep}{-1pt}  
\begin{deluxetable*}{cccccccccccccccccccc}
\tablecaption{Number (fraction) of S0 galaxies harboring rings in each type of spectral map according to activity and quenching classes\label{tab:ring_numbers}.}
\tablewidth{0pt}
\tablehead{
&  & \multicolumn{6}{c}{EW(\Ha)} & \multicolumn{6}{c}{$L^\prime$(\Ha)} & \multicolumn{6}{c}{\Db} \\ \hline
Spectral & Quenching & \multicolumn{2}{c}{Ringed} & \multicolumn{2}{c}{Doubtful} & \multicolumn{2}{c}{Non-ringed} & \multicolumn{2}{c}{Ringed} & \multicolumn{2}{c}{Doubtful} & \multicolumn{2}{c}{Non-ringed} & \multicolumn{2}{c}{Ringed} & \multicolumn{2}{c}{Doubtful} & \multicolumn{2}{c}{Non-ringed} \\
Class & Class & & & & & & & & & 
}
\startdata
\multirow{4}{*}{PS} & IO & 20 & (0.56) & 10 & (0.28) & 6 & (0.17) & 15 & (0.56) & 4 & (0.15) & 8 & (0.30) & 3 & (0.12) & 14 & (0.54) & 9 & (0.35) \\
 & OI & 0 & (0.0) & 0 & (0.0) & 3 & (1.0) & 0 & (0.0) & 0 & (0.0) & 1 & (1.0) & 0 & (0.0) & 0 & (0.0) & 1 & (1.0) \\
 & Flat & 2 & (0.05) & 2 & (0.05) & 34 & (0.89) & 1 & (0.02) & 0 & (0.0) & 42 & (0.98) & 0 & (0.0) & 1 & (0.01) & 108 & (0.99) \\
 & All & 22 & (0.29) & 12 & (0.16) & 43 & (0.56) & 16 & (0.23) & 4 & (0.06) & 51 & (0.72) & 3 & (0.02) & 15 & (0.11) & 118 & (0.87) \\ \hline
\multirow{4}{*}{AC} & IO & 14 & (0.48) & 11 & (0.38) & 4 & (0.14) & 0 & (0.0) & 4 & (0.15) & 23 & (0.85) & 4 & (0.16) & 9 & (0.36) & 12 & (0.48) \\
 & OI & 1 & (0.01) & 1 & (0.01) & 65 & (0.97) & 0 & (0.0) & 0 & (0.0) & 62 & (1.0) & 0 & (0.0) & 1 & (0.02) & 60 & (0.98) \\
 & Flat & 0 & (0.0) & 1 & (0.14) & 6 & (0.86) & 0 & (0.0) & 0 & (0.0) & 8 & (1.0) & 0 & (0.0) & 0 & (0.0) & 9 & (1.0) \\
 & All & 15 & (0.15) & 13 & (0.13) & 75 & (0.73) & 0 & (0.0) & 4 & (0.04) & 93 & (0.96) & 4 & (0.04) & 10 & (0.11) & 81 & (0.85) \\ \hline
All & All & 37 & (0.21) & 25 & (0.14) & 118 & (0.66) & 16 & (0.10) & 8 & (0.05) & 144 & (0.86) & 7 & (0.03) & 25 & (0.11) & 199 & (0.86)
\enddata
\tablecomments{Limited to galaxies with vectorizable spectral profiles, non-extreme inclinations ($i\leq 60^\circ$) and robust to PSF effects (i.e., $\sigma_{\rm PSF}/R_{e} < 0.6$).}
\end{deluxetable*}
\addtolength{\tabcolsep}{1pt}

Visual identification of the dominant trends in the spectral maps unveils the existence of a pattern that, although not overwhelmingly prevalent, is quite widespread: the presence of ring-shaped inner structures ($0.5 R_\mathrm{e}\lesssim R\lesssim 1.5 R_\mathrm{e}$) in EW(\Ha). These spectral rings are also echoed sometimes in the spatial distributions of $L^\prime$(\Ha) and \Db\, as well as in the corresponding optical images, although in this latter case with a significantly lower contrast that makes their detection more difficult. To avoid personal bias in ring identification, all the authors of this paper have participated in the visual inspection of the MaNGA spectral maps, after adjusting them to comparable angular sizes. We have followed a scheme in which each of us has independently assigned an integer score from $0$ to $2$ to the maps depending on whether the presence of a ring is rejected (0), confirmed (2), or doubtful (1), and added the individual scores afterwards. Spectral images are divided into three categories based on the overall ring score ($r$) they have attained: ringed ($r \ge 6$), doubtful ($r=3$--$5$), and non-ringed ($r < 3$). Furthermore, to obtain a reliable estimate of the ring abundance in our S0 sample, we have restricted the search to galaxies not highly inclined ($i < 60^\circ$), with linear PC1--PC2 radial profiles, and with a good seeing, a condition that we define as $\sigma_{\rm PSF}/R_\mathrm{e} < 0.6$ after verifying that the removal of galaxies that are small relative to the PSF according to this constraint does not introduce spurious trends either in ring detection or as a function of galaxy size. These additional cuts yield a subset of 275 S0s with essentially the same distributions of effective radii and stellar masses as its parent sample shown by the green histograms of Figure~\ref{fig:sample_properties}, as well as similar fractions of spectral and quenching classes. Still, not all spectral maps in this final high-quality subset show sufficiently continuous spatial coverage to assess the presence of rings. The figures in Table~\ref{tab:ring_numbers} give detailed information about the number of maps of each type that have finally been used.
\begin{figure*}[ht!]
\gridline{\fig{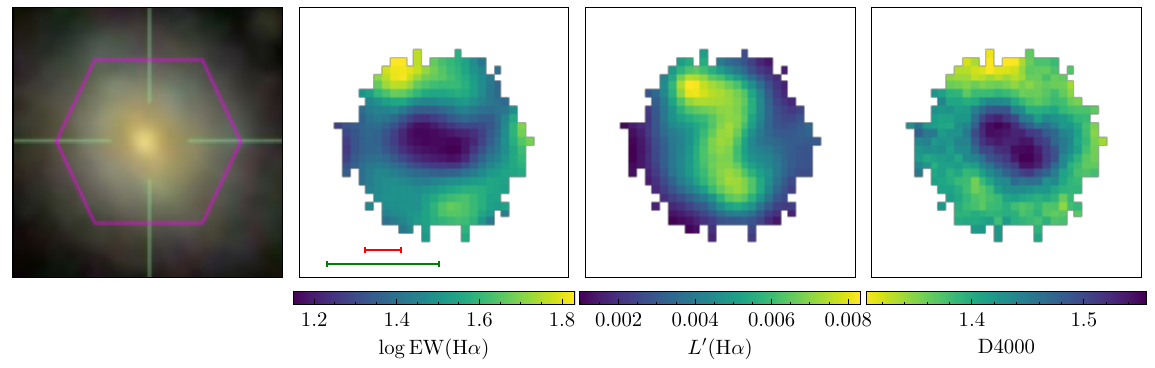}{0.95\textwidth}{}}
\vspace*{-1.6\baselineskip}
\caption{From left to right, optical image and EW(\Ha), $L^\prime$(\Ha) and \Db\ spectral maps for the IO-quenched S0--AC $9506$--$1901$. The green and red horizontal lines included in the EW(\Ha) panel indicate, respectively, the galaxies' Petrosian half-light diameter in the $r$ band and $2\,\sigma_{\rm PSF}$. Note what appears to be a barred spiral-shaped feature outlined by the dense gas distribution on the \Ha\ luminosity map, which, however, is not evident in the optical image nor in the other two spectral maps. Montages like this one for all the vectorizable galaxies in the MaNGA S0 sample are shown in Appendix~\ref{A:all_maps}. The ring score of this galaxy in the EW(\Ha) is $r = 3$.}
\label{fig:hidden_spiral}
\end{figure*}

Applying this strategy, we find that the global fraction of nearby S0s with a ring in the EW(\Ha) is 0.21, which rises to 0.34 if we include those galaxies for which ring detection has been deemed inconclusive (i.e., with $r\geq 3$). These values are substantially larger than the occurrence rate of $\sim 7\%$ inferred by \citet{Deeley+2021} from both simulations and observations. However, we note that in their case, the detection of ringed galaxies was carried out using a single late snapshot from a cosmological run and, observationally, the \Ha\ fluxes. In fact, the near $10\%$ fraction of rings with $r\geq 6$ that we infer from our $L^\prime$(\Ha) maps does not stray notably from \citeaposauthor{Deeley+2021}. As shown in Table~\ref{tab:ring_numbers}, EW(\Ha) rings are more frequent among the members of the PS class than in S0--ACs, with respective fractions of $\sim 0.29$ and 0.15, assuming again only clear detections. Besides, the mean value of EW(\Ha) in the spaxels defining these annular structures in AC systems is about four times larger than in objects of the PS class (we discuss the implications in \S~\ref{S:scenario}). We also find that essentially all S0s with a ring in this spectral index have IO quenching, while more than half of the galaxies with this kind of spectral profile are ringed ($85\%$ if we include objects with doubtful rings). In contrast, spectral rings are infrequent in S0 galaxies with either OI-quenched or flat spectral profiles. 

Examples of S0 galaxies showing different combinations of quenching and activity classes are provided in Figure~\ref{fig:rings}. As can be seen from this plot and Table~\ref{tab:ring_numbers}, the ringed structures are usually more conspicuous in the EW(\Ha) maps than in the maps of the other two spectral indices. Examination of the montages of optical images and spectral maps provided in Appendix~\ref{A:all_maps} shows that most of the resolved $L^\prime$(\Ha) distributions have in common the presence of a peak in the gas density that encompasses the most central spaxels. Nonetheless, in a good number of instances, the values of EW(\Ha) achieved in these nuclear gaseous enhancements indicate that they are not forming new stars, but possibly tracing low-ionization emission-line regions fueled by old post-AGB stars (see also, e.g., Fig.~\ref{fig:rings}, column (c)). On the other hand, few of the rings with enhanced SSFR detected in the IO-quenched S0--ACs are visible in the associated $L^\prime$(\Ha) maps. In fact, in galaxies of this type, the radial extent of the cold gas tends to lie within the outer boundary of the rings, indicating that the distribution of ionized hydrogen in this area has a strong negative radial gradient (see, e.g., Fig.~\ref{fig:rings}, column (d)). In contrast, the stellar populations, as traced by \Db, describe patterns very similar to those found in the EW(\Ha) maps, with younger stars concentrating at the location of the SF rings. 

\emph{The case of \object{MaNGA 9506--1901}.} Among all the S0s investigated, this object has drawn our attention because its $L^\prime$(\Ha) map shows what resembles a double spiral arm-like feature of ionized hydrogen connected by a similarly intense bar of warm gas. As shown in Figure~\ref{fig:hidden_spiral}, this is a nearly face-on S0--AC with a global IO-quenched spectral profile that concentrates the most important star formation activity at the ends of the gaseous spiral arms, where the EW(\Ha) index reaches peak values that exceed 40 \AA. Inspection of the \Db\ map suggests that the nebular spiral structure, assuming it behaves as a single body, may be winding in a leading sense. Interestingly, there is no visible sign of the presence of spiral arms or bars in the stellar component of the galaxy in any of the SDSS filters, not even in deeper optical images coming from the Dark Energy Camera Legacy Survey \citep{DESI+19}. Only a faint bluish band is perceived in the optical image along the interior of the IFU's hexagonal boundary that could be spatially correlated with the peripheral zones of enhanced SSFR and younger stellar ages that can be seen, in green/yellow colors, in the corresponding panels of Figure~\ref{fig:hidden_spiral}. On the other hand, the fact that the central bar-like section of this spectral feature lacks a visible counterpart reinforces the idea that it is a purely gaseous structure not associated with star formation. Early-type galaxies such as this one with eye-catching features of ionized hydrogen have also been documented in other IFS surveys like CALIFA \citep{Gom+16}. They call for follow-up observations that may end up providing valuable insights for advancing in our understanding of the buildup history of these systems.

\section{Ring fraction versus global properties}
\label{S:origin_of_rings}

The next step is to analyze the behavior of the fraction of EW(\Ha) rings in relation to two global properties of the studied systems ---one internal, the stellar mass, and the other external, the environmental density--- that bear a direct relationship with their formation channels. We aim at gathering evidence that helps to unveil the role that such fundamental properties play in shaping the rings and to establish later on the relationship that these transient structures may have with the evolutionary histories of the S0 galaxies harboring them ---which, as we have seen in the previous section, are almost exclusively those with IO-quenched spectral profiles.

\begin{figure*}[ht!]
\vspace*{0.0\baselineskip}
\gridline{\fig{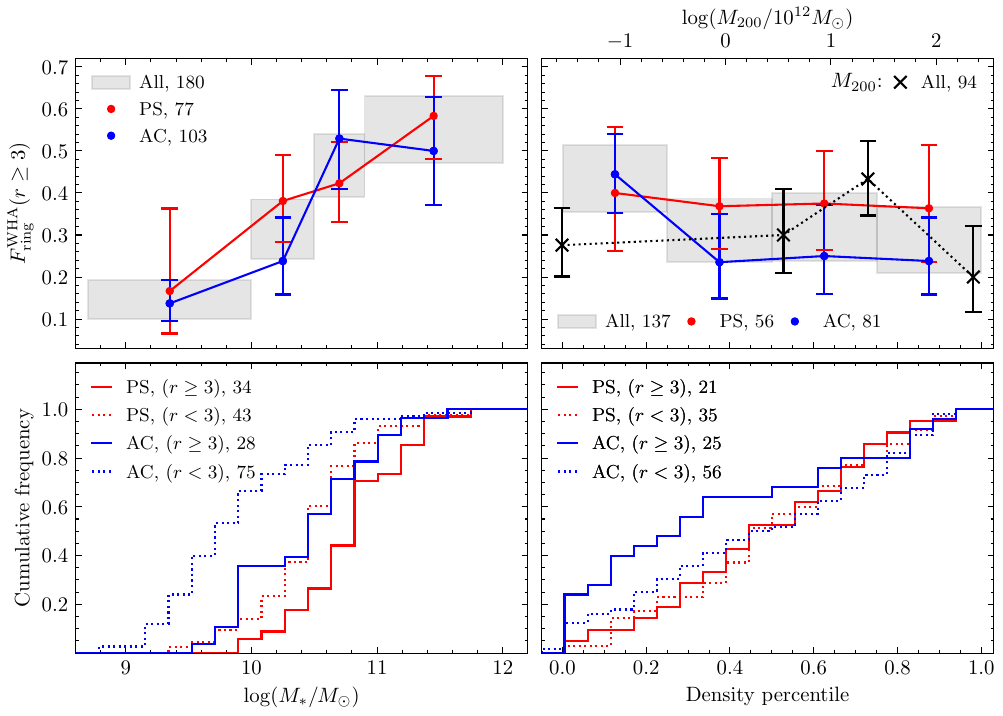}{0.97\textwidth}{}}
\vspace*{-1.5\baselineskip}
\caption{Fractions (top) and cumulative frequencies (bottom) of varying subsets of present-day S0s harboring or likely harboring EW(\Ha) rings as a function of their total stellar mass (left) and environment (right). The latter is quantified in the upper panel using two different measures: a ranked density percentile inferred from a nearest-neighbor estimator (colored circles, lower $x$-axis) and the virial mass $M_\mathrm{200}$ of the halo hosting the galaxies (black crosses, upper $x$-axis). The error bars on the fractions show $1\sigma$ uncertainties estimated by adopting a Wilson score interval approximation to the binomial error distribution. The number of galaxies in each subset is provided.}
\label{fig:comparisons}
\end{figure*}

As shown in the top left panel of Figure~\ref{fig:comparisons}, the fraction of galaxies in our high-quality S0 sample that harbor or likely harbor EW(\Ha) rings, $\Fring (r\geq 3)$, is positively correlated with the measurement of the total stellar mass. It rises monotonically from $\sim 10\%$--$20\%$ among the S0s in the smallest mass bin ($M_\ast\leq 10^{10}\,M_\odot$) to $\sim 50\%$--$60\%$ for objects in the more massive one ($M_\ast\geq 10^{11}\,M_\odot$). If the data are partitioned by activity class, this behavior is still maintained. If anything, it might be inferred that the ring fraction for S0--ACs peaks at $M_\ast\sim 10^{10.5}$--$10^{11}\,M_\odot$. Note, however, that while we have verified using different binning strategies that the observed positive correlation is robust, the quoted fractional abundances of rings are relatively uncertain given the limited size of the subsamples employed (listed in the legend of each panel of the figure). The cumulative frequency plot in the bottom left panel of Figure~\ref{fig:comparisons} ---which allows one to compare the distributions of the number of observations as a function of the stellar mass in a binning-independent manner--- confirms that ringed S0 galaxies tend to be more massive than their non-ringed counterparts, regardless of their spectral class. A two-sample Kolmogorov--Smirnov (K-S) test rejects the null hypothesis that ringed and non-ringed subsets arise from the same parent distribution at level $\alpha=0.01$ for both the PS and AC classes ($p$-values of 0.01 and 0.0007, respectively). These findings are in line with recent work by \citet{DiaG+19}, who report, from a subset of the S$^4$G mid-IR sample \citep{She+10} of nearby disks ($d\lesssim 40$ Mpc) encompassing a similar range of masses, that SF rings are preferentially hosted by massive, gas-poor, centrally concentrated galaxies that are not dominated by dark matter in their innermost regions (see also \S~\ref{S:scenario}). 

The influence of the environment is evaluated by means of our own nearest-neighbor (NN) estimator of the local density. It is a parameter based on the distances up to the fifth neighbor measured using a Bayesian metric (see eqs.\ [1] and [2] in  \citetalias{Tous+2020}), which is then ranked according to the percentile that each galaxy gets in the deepest volume-limited subset that can be defined for it. Prior to running this test, however, we have examined the extent to which its results could be influenced by the observed positive correlation of the ring fraction with stellar mass. With this purpose, we have split our sample into four evenly spaced density percentile intervals centered at 0.125, 0.375, 0.625, and 0.875, and compared the distributions of the stellar mass in them, whose means and standard deviations have turned out to be $10.23\pm 0.52$, $10.39\pm 0.47$, $10.29\pm 0.63$, and $10.29\pm 0.61$, respectively. The fact that these values are broadly consistent with each other implies that, at least for the present subset of observations, stellar mass does not correlate with environmental density, and therefore that any trend shown by the fraction of rings as a function of the latter will not be a consequence of its underlying relationship to stellar mass.

As can be seen in the top right panel of Figure~\ref{fig:comparisons}, the abundance of rings remains constant, within the calculated uncertainties, in the four equal-sized percentile intervals defined above. This result, which is maintained when the PS and AC activity subsets are analyzed separately, also agrees with the findings of \citet{DiaG+19}, who claim that the galaxies' environs do not play a relevant role in ring formation. To determine the extent to which the procedure applied to probe the environment could be conditioning the outcomes ---\citeauthor{DiaG+19} also use a local NN estimator but relying on three neighbors--- we have reevaluated the ring fractions using a more global estimator of environmental density, namely the virial mass, $M_\mathrm{200}$, of the host group of the galaxies calculated by \citet{Tempel+2017} from the observed velocity dispersion and group extent on the sky (see their eq.\ [3]), which is available for 94 of the S0s in the high-quality subset. Using this new parameter, after certifying again that our data do not hide a relationship between stellar mass and environment, we still find the ring fraction to be essentially unrelated to the characteristics of the region in which the galaxies reside (crosses in the top right panel of Fig.~\ref{fig:comparisons}). Likewise, the associated cumulative distributions of counts of ringed and non-ringed galaxies with environmental density do not show statistically significant differences, as reflected by the high $p$-values of 0.996 (PS) and 0.226 (AC) delivered by the K-S test. However, in fairness, we acknowledge that the robustness of all these results is not foolproof, since they are based on a relatively limited number of measurements. Relationships such as those investigated in this section and their dependence on the spectral classification of galaxies will need to be corroborated with the help of datasets that are not only much larger than currently available samples, but that are also highly complete, free from aperture effects, and capable of providing uniform coverage across the full range of stellar environments and masses.

\section{A scenario for the formation of SF rings in S0 galaxies}
\label{S:scenario}

As commented in the Introduction, star formation in present-day S0 galaxies is observed to be frequently organized in ring-like structures \citep[e.g.][]{Sal+12}. According to the investigation of the formation pathways of S0s by \citet{Deeley+2021} from the data generated in the IllustrisTNG-100-1 cosmological simulation \citep{Spr+18}, practically all galaxies of this type experience at some point in their lives a transitional phase in which they host an SF ring. In the simulation, rings always appear related to merger episodes, thereby showing that this mechanism is their major driver in front of other feasible formation channels, such as the inflow of fresh gas through filaments or outflows from active galactic nuclei. \citeauthor{Deeley+2021} also observe that the onset of SF rings is preceded by the depletion of the gas in the central region of the hosts, and often accompanied by the formation of a stellar bar ---or other non-axisymmetric distortion of the gravitational potential set up by the merger itself \citetext{see \citealt{Com+14} and references therein}, which indicates that these phenomena are linked for the most part to disk instabilities of the primary galaxy. This agrees with our finding that essentially all our S0s with clear or tentative spectral rings, whether of the AC or PS class, exhibit IO quenching. In this sense, it is hard not to notice that some of our $L^\prime$(\Ha) maps of ringed, IO-quenched S0s (see, for example, the middle panel in Fig.~\ref{fig:rings}c) bear a close resemblance, if one takes into account the distorting effects of the PSF, with the gas density distribution associated with the coplanar ring formed in the hydrodynamic simulations of mergers of a lenticular galaxy with a small gas-rich satellite by Mapelli et al. (\citeyear{Mapelli+2015}, see their Figure~4). The agreement with this latter work also extends to quantitative results, since our finding that EW(\Ha) rings are about twice as frequent in PS sources as among AC ones is in good agreement with the shorter lifetimes shown by the simulated gaseous rings in regions of intense star formation because the thermal energy released by supernovae explosions easily destroys them. However, unlike what \citeauthor{Mapelli+2015} suggest with their simulations, the fact that we observe that the \Ha\ emission from the rings of the S0--ACs is about four times higher than that recorded in the rings of S0 categorized as PS (\S~\ref{S:rings}), implies, provided the gas content of the merged satellites does not correlate with the spectral classification of the central objects, that star formation in these structures is largely fueled by gas remaining in the primary galaxies rather than by gas supplied by the satellites, especially for the S0--ACs. In turn, this result suggests that the \Ha\ rings are very possibly coplanar with the disks and it is therefore consistent with the widely accepted standard theory of ring formation by gas accumulation at bar resonances.

The rest of the results allow us to infer the specifics of the proposed merger scenario for the formation of SF rings in S0 galaxies by showing that such annular structures are more likely to arise in mergers involving a massive primary galaxy closely orbited by a tiny dwarf satellite. It is well known both through simulations \citep[e.g.][]{VSH10} and observations \citetext{e.g.\ \citealt{Mas+12}; see also \citealt{ZMW21} and references therein} that the presence of a dissipative gaseous component prevents the formation of disk instabilities, bars, and/or rings, by exchanging energy and angular momentum with the stars. Therefore, it does not seem to be coincidental our finding that the ring fraction increases with the mass of S0 galaxies, reaching its peak around $M_\ast\sim 10^{10.5}$--$10^{11}$\,\Msun, a mass scale of the order of the golden mass. The latter is a well-constrained, redshift-independent quantity inferred in theoretical studies of galaxy formation \citetext{\citealt{DB06}; see also \citealt{DLD19} and \citealt{Dek+21}} ---it corresponds to a virial halo mass $M_\mathrm{vir}\sim 10^{12}$\,\Msun--- that marks a general bimodality in many galaxy properties. In particular, the efficiency of star formation increases progressively as the mass of galaxies approaches this characteristic value, causing the gaseous-to-stellar mass ratio to reach a minimum for golden-mass objects. This would explain the observed tendency of ringed S0s to be more massive than their non-ringed counterparts.

On the other hand, a major merger or a minor one involving a relatively massive satellite could transfer more energy than is required to create a transient instability in a disk marginally stable against developing it. Any resonance that cannot be efficiently damped by the dissipative gaseous component will grow out of control and eventually self-destruct, whereas if the mergers are excessively disruptive, no resonances will be created at all. Furthermore, in a typical merger scenario, the fraction of \Ha\ rings, provided these structures do not last very long, would be expected to increase with increasing environmental density, peaking at roughly the scales of groups, such as happens with the merger rate efficiency, and then decrease its value in the most overdense regions. However, we have reported evidence for the lack of correlation between the frequency of rings and different metrics that probe the density of the environment. So, for this result to support the picture being built here, it is necessary to assume that the role of satellite is played mostly by tiny dwarf galaxies orbiting in the innermost regions of the halos of the primary galaxies. This should allow the two progenitor galaxies to retain their merging pair status even if the system falls into a densely populated region, and therefore the merger rate efficiency to remain aloof from the physical conditions dictated by a changing local environment. Interestingly, in the example of a typical spiral galaxy transforming into an S0 through multiple merger events shown in the cosmological simulations analyzed by \citet{Deeley+2021}, it is the merger that takes place at $\sim 7$~Ga, involving what is, by far, the smallest satellite that participates in the buildup of this galaxy, the one that triggers the formation of a distinct SF ring (see their Figures~9 (a) and (e)). Although we lack enough information to rigorously constrain the mass ratio defining these mini-mergers, we venture to speculate, exclusively on the basis of simulations such as those of \citet{Mapelli+2015}, that it could well be significantly less than 1:10 and involve star-forming dwarf-galaxy satellites with gas masses of a few times $10^8\;$\Msun, comparable to those of the Large and Small Magellanic Clouds, which are in the process of falling toward our Galaxy \citep{Bes15}.

\section{Summary and conclusion}
\label{S:conclusions}

We have studied the spatially resolved optical spectral properties of a sample of 532 lenticular galaxies from the MaNGA survey. The spaxels of the data cubes, once properly corrected and cleaned, have been stacked in radial bins and projected in the plane defined by the first two principal components resulting from the principal component analysis of their single-fiber spectra. The information has been further synthesized by best-fitting these spectral radial profiles by means of straight lines that have been then converted into vectors in the PC1--PC2 plane.

This dimensionality reduction process has allowed us to show that present-day S0s can be classified into three groups according to the sign of the inner radial gradients ($R\lesssim 1.5\,R_\mathrm{e}$) of their star formation: galaxies with inside-out quenched radial profiles, systems with outside-in quenching; and galaxies with flat gradients. Regarding the global star formation activity, we have used the \citet{Tous+2020} PCA-based diagnostic to divide the galaxies into two main categories: passive sequence and active cloud, encompassing galaxies whose single-fiber spectra are representative, respectively, of passive and active systems, and including in the latter class a few objects with intermediate spectral characteristics assigned by these authors to the transition region.

Visual inspection of the EW(\Ha) maps of these galaxies has revealed the existence of a pretty significant fraction of objects hosting inner rings ---$34\%$ if dubious ringed systems are included--- which, however, are generally absent from maps constructed from other spectral indicators of stellar age such as $L$(\Ha) and the \Db\ index, as well as in the optical SDSS images. The preferential detection of SF rings by their EW(\Ha) emanates not only from the direct relationship between the strength of the \Ha\ emission and the SSFR, but also from the fact the EW provides information on the intensity of the lines normalized to the underlying continuum, thus increasing their contrast in spectral maps. The present sample of S0s constitutes, in fact, the largest collection of ringed galaxies ever identified in the local universe through spatially resolved spectroscopic observations.

The analysis of the relationships of the spectral rings to the physical properties of their host S0 galaxies has produced the following results. 

\begin{enumerate}    
    \item In almost all cases, the SF rings are associated with IO-quenched radial profiles of star formation (Table \ref{tab:ring_numbers}). This is in agreement with state-of-the-art cosmological simulations, which show that rings of this type are always related to merger episodes and that their onset is preceded by the depression in the level of star formation in the central region of their host galaxies and frequently linked to other disk instabilities such as bars.

    \item The \Ha\ emission in the rings of the S0--ACs is about four times greater than that of the members of the PS class. This suggests that star formation in these rings, especially for objects of the former class, is fueled primarily by gas that still remains in the host galaxies rather than gas provided by the merging satellites. This also suggests that the detected spectral rings are probably mostly coplanar structures.
    
    \item The EW(\Ha) rings are found to be about twice as frequent in PS systems as in AC ones (Table \ref{tab:ring_numbers}). This is in line with controlled simulations of ring formation in lenticular galaxies via mergers of small satellites, which show that the more intense the star formation, the more easily these transient structures are destroyed by supernova feedback.
    
    \item Whatever the activity class of the S0s, their ring fraction increases with stellar mass, showing a tentative peak around a mass scale $M_\ast\sim 10^{10.5}$--$10^{11}\,M_\odot$ of the order of the (golden) mass that marks the maximum of star formation efficiency and the low point of the gaseous-to-stellar mass ratio in galaxies (Fig. \ref{fig:comparisons}, left-hand panels). Such behavior, which has also been observed in the general galaxy population, can be explained by the well-known fact that disk instabilities have more difficulties forming in the presence of large amounts of gas than in gas-poor systems.
    
    \item There is no statistically significant evidence that the frequency of \Ha\ rings in S0 galaxies correlates with environmental density, which we have measured using two different standard metrics (Fig. \ref{fig:comparisons}, right-hand panels). This suggests that the formation of SF rings in lenticular galaxies is not ruled by their environment.

\end{enumerate}
 
Based on these findings, and on the physical conditions necessary for the formation of in-plane SF rings in galaxy disks, we conclude that the presence of annular structures in the EW of the \Ha\ emission line among fully formed S0s is a relatively common phenomenon, possibly originating from disk resonances driven by weakly disruptive mini-mergers, and preferentially involving a relatively massive primary galaxy and a tiny companion strongly bound to the former.

We leave for a future article (Tous et al., in preparation) a detailed discussion of how this scenario for ring formation fits into a more extensive and complex picture describing the multiple formation channels that S0 galaxies seem to follow.

\section*{Acknowledgements}
We acknowledge financial support from the Spanish state agency MCIN/AEI/10.13039/501100011033 and by `ERDF A way of making Europe' funds through research grants PID2019--106027GB--C41 and PID2019--106027GB--C43. MCIN/AEI/10.13039/501100011033 has also provided additional support through the Centre of Excellence Severo Ochoa's award for the Instituto de Astrof\'\i sica de Andaluc\'\i a under contract SEV--2017--0709 and the Centre of Excellence Mar\'\i a de Maeztu's award for the Institut de Ci\`encies del Cosmos at the Universitat de Barcelona under contract CEX2019--000918--M. J.L.T.\ acknowledges support by the PRE2020--091838 grant from MCIN/AEI/10.13039/501100011033 and by `FSE Invests in your future'. H.D.S.\ acknowledges support by the PID2020-115098RJ-I00 grant from MCIN/AEI/10.13039/501100011033. The authors also wish to thank an anonymous referee for comments and suggestions that have helped to improve the presentation of the results.

\bibliography{biblio}{}
\bibliographystyle{aasjournal}

\appendix

\section{Classification of S0 galaxies from the principal component analysis of their spectra}
\label{A:PCA}
\renewcommand{\thefigure}{A.\arabic{figure}}
\setcounter{figure}{0}

In this work, we use the first two eigenspectra derived in the principal component analysis (PCA) of a sample of $68,043$ extinction-corrected single-fiber SDSS optical spectra (\citetalias{Tous+2020}; see also \citealt{JimP+22}). Here we provide a brief outline of the method applied and the resulting spectral classes.

A dataset of $N$ SDSS galaxy spectra can be thought of as a set of flux vectors, $f_i$, in a space of $\sim 3800$ dimensions or (vacuum) wavelength intervals. Each one of these vectors can be expressed exactly as the sum of a vector of mean fluxes, $\langle f\rangle$, and a linear combination of $M$ orthonormal vectors or eigenspectra, $e_j$, that account for all the variability around the mean, in the form \citep[e.g.][]{Connolly+1995}
\begin{equation}
    f_i = \langle f\rangle + \sum_{j=1}^{M} \mbox{PC}j_i\ e_j\ \ \ \ \ \ \ \ \ \ \ 1\leq i\le N\;,
    \label{eq:flux_approximation}
\end{equation}
where the coefficients PC$j_i\equiv f_i\cdot e_j$ are the projections of each individual spectrum on the new base of principal components arranged in decreasing order of their relative power, i.e., of the proportion of the total variance of the data that they explain. Thus, when the main variance of a data set lies in a low-dimensional space, one can get a good visualization of it by truncating the above expansion to the first few eigenvectors. As we showed in \citetalias{Tous+2020}, once any extrinsic source of variability in the spectra is removed, nearly $90\%$ of the variance in the SDSS sample of S0s lies in the subspace whose axes are the first two principal components, a huge dimensional reduction with a modest loss of information that allows one to carry out the spectral classification of these objects. Furthermore, as we will show in Tous et al., in preparation, the first principal components resulting from the PCA of the composite spectra inferred from the MaNGA data cubes are essentially identical to those obtained from the single-fiber spectra. This implies that the spatially resolved spectra have the same main sources of intrinsic variability and can therefore also be classified using the same 2D subspace. The mean optical spectrum of the S0s and the first five eigenspectra inferred from the SDSS data are available in \citet{JimP+22_table}.

\begin{figure}[htb!]
\gridline{\fig{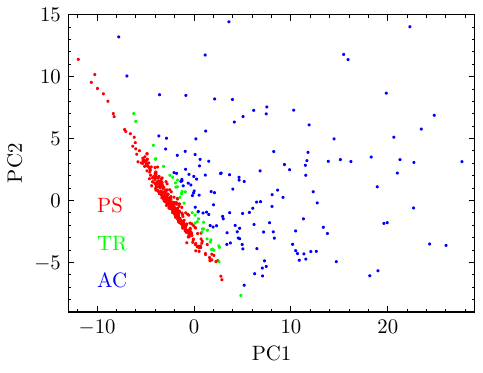}{0.72\columnwidth}{}}\vspace*{-2.5\baselineskip}
\caption{Projections in the PC1--PC2 subspace of the single-fiber SDSS optical spectra of the 532 bona fide galaxies included in our main MaNGA S0 sample. The diagram is subdivided into three distinct regions corresponding to three different spectral classes: passive sequence (PS, red), transition region (TR, green), and active cloud (AC, blue).}
\label{fig:pc1_pc2}
\end{figure}

The classification of lenticular galaxies within the PC1--PC2 plane in \citetalias{Tous+2020} revealed the existence of two clearly distinct regions outlined by subpopulations of S0s with statistically inconsistent physical properties: a very compact and densely populated narrow band that diagonally crosses the PC1--PC2 subspace and a less crowded and much more scattered zone running from the right of it (see also Fig.~\ref{fig:pc1_pc2}). These regions were identified as the `passive sequence' (PS) and the `active cloud' (AC), because they encompass sources representative, respectively, of passive and active sources. Compared to the S0--PS, which have absorption-dominated spectra, the S0--AC exhibit spectra with significant nebular emission, and on average they are also somewhat less massive, are more luminous but with less concentrated light profiles, possess a younger, bluer, and metal-poorer stellar component, and avoid environments of high galaxy density. These two main areas are separated by a narrow dividing zone, dubbed the `transition region' (TR), encompassing objects with intermediate spectral and physical characteristics. This completes a spectral classification for the S0s of the local universe that is reminiscent of the well-known `red sequence-green valley-blue cloud' division applied to the entire galaxy population in color-magnitude diagrams (see \S~5.2 in \citetalias{Tous+2020} for details).

In Figure~\ref{fig:pc1_pc2}, we show the projections in the PC1--PC2 subspace of the SDSS optical spectra of the 532 bona fide S0 galaxies that constitute our MaNGA sample. Solid red dots are used to identify PS members, which encompass $64\%$ of the galaxies in the sample, blue dots are for the $29\%$ of the S0s that belong to the AC class, while the remaining $7\%$ of S0s that fall in the TR are drawn using green colored dots (see also \S~\ref{S:data}).

\newpage
\renewcommand{\thefigure}{B.\arabic{figure}}
\setcounter{figure}{0}
\section{Optical images and spectral maps of S0 galaxies with linear PC1--PC2 radial profiles}
\label{A:all_maps}
In this appendix we show the montages of optical images and spectral maps for S0 galaxies with linear PC1–PC2 radial profiles, emphasizing those objects harboring an unequivocal ring in the EW(\Ha) map. The galaxies appear sorted by quenching type (inside out, IO; outside in, OI; and flat, F) and spectral class (passive sequence, PS; active cloud, AC; and transition region, TR).
\begin{figure}[hbt!]
\gridline{\fig{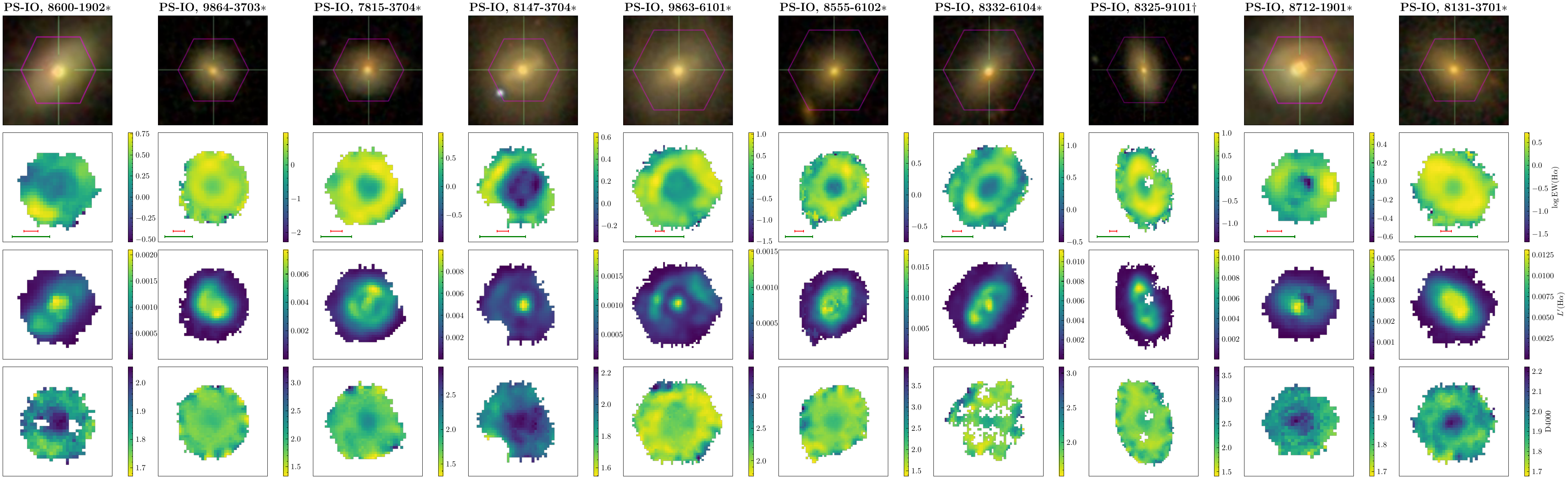}{\textwidth}{}}\vspace*{-2.1\baselineskip}
\caption{Montages of the optical images and the three spectral maps for all the S0 galaxies with linear PC1–PC2 profiles. The green and red horizontal lines included in the EW(\Ha) panels indicate, respectively, the galaxies’ Petrosian half-light diameter in the $r$ band and $2\,\sigma_{\rm PSF}$. An asterisk next to each galaxy’s identifier indicates the presence of an unequivocal ring in the EW(\Ha) map ($r \geq 6$), unless the galaxy is highly inclined ($i > 60^\circ$) or has a poor seeing condition ($\sigma_{\rm PSF}/R_{e} \geq 0.6$), in which case a dagger is used.}
\label{fig:set_maps}
\end{figure}

\setcounter{figure}{0}
\begin{figure}[htb!]
\gridline{\fig{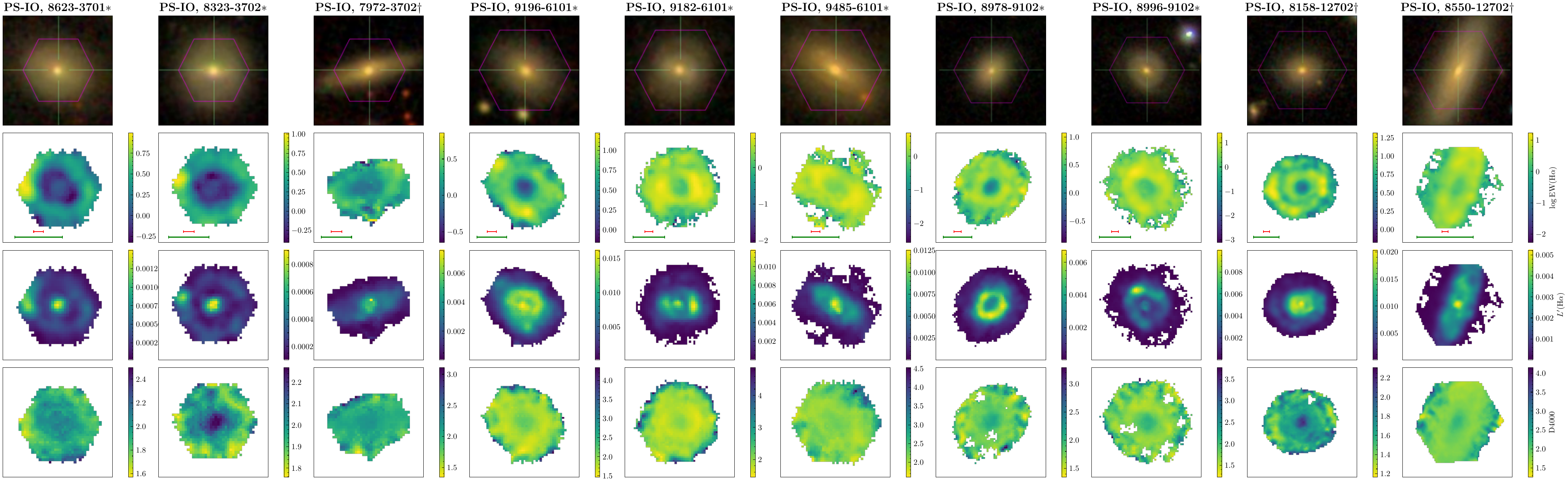}{\textwidth}{}}\vspace*{-2.5\baselineskip}
\gridline{\fig{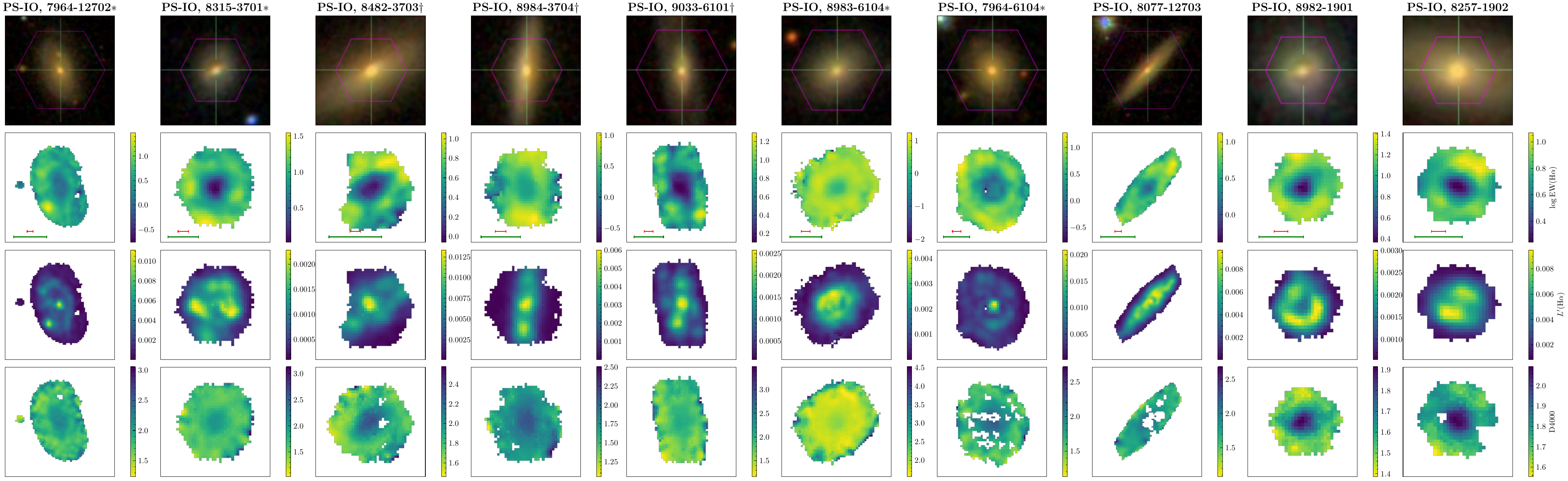}{\textwidth}{}}\vspace*{-2.5\baselineskip}
\gridline{\fig{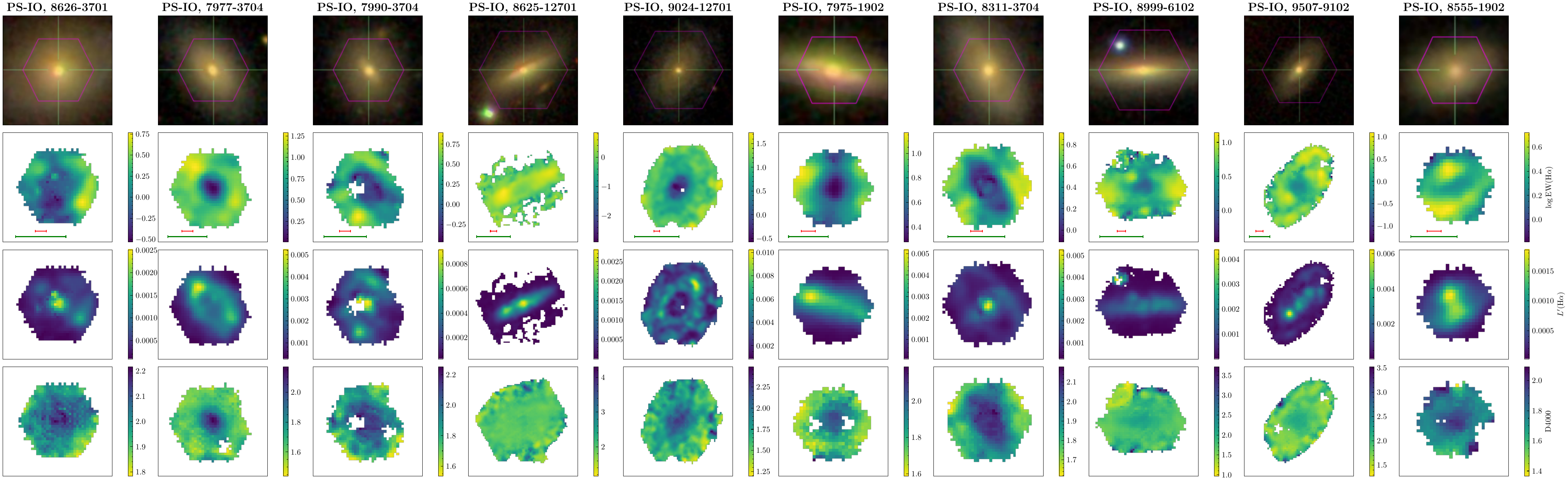}{\textwidth}{}}\vspace*{-2.5\baselineskip}
\gridline{\fig{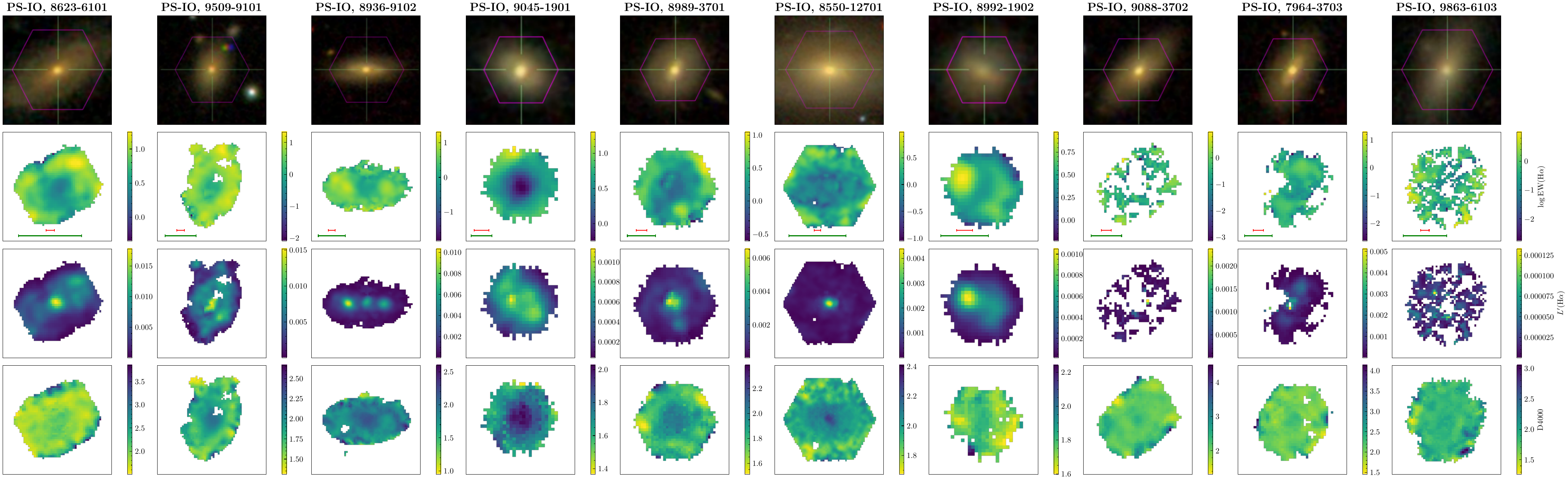}{\textwidth}{}}\vspace*{-2.5\baselineskip}
\caption{Cont.}
\end{figure}

\setcounter{figure}{0}
\begin{figure}[htb!]
\gridline{\fig{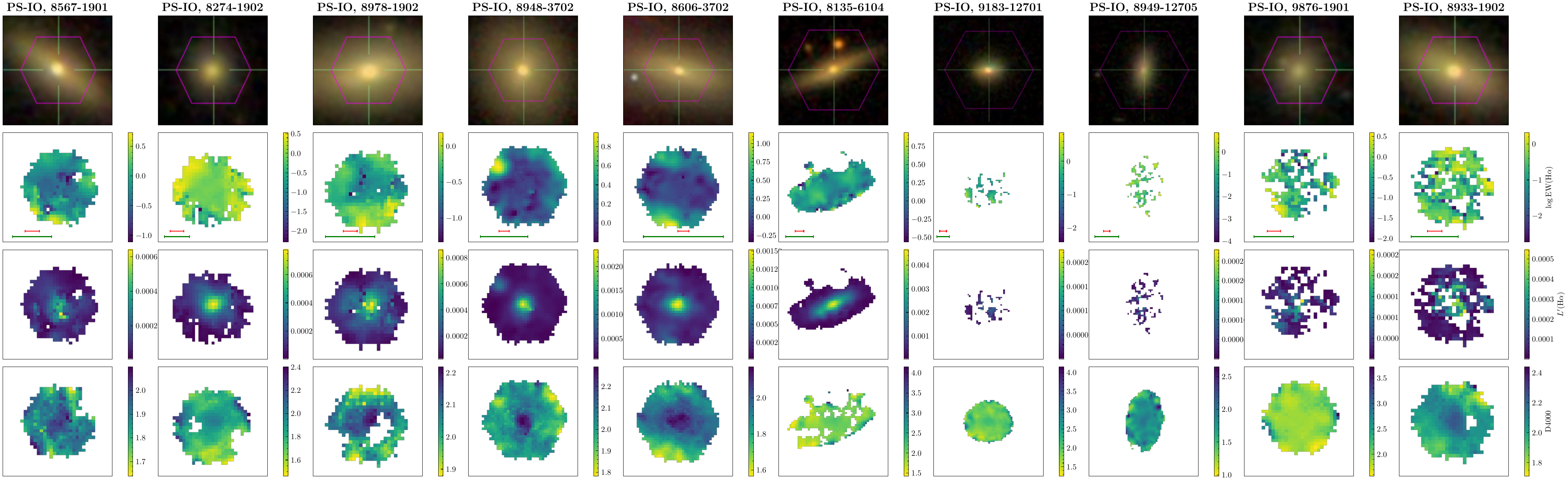}{\textwidth}{}}\vspace*{-2.5\baselineskip}
\gridline{\fig{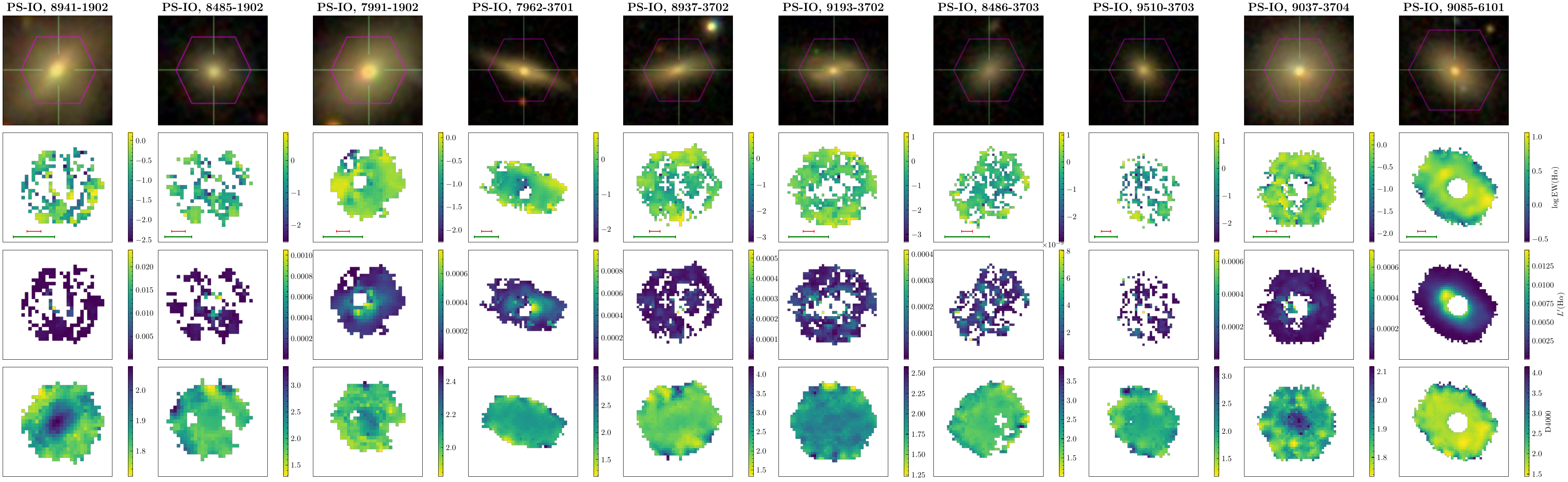}{\textwidth}{}}\vspace*{-2.5\baselineskip}
\gridline{\fig{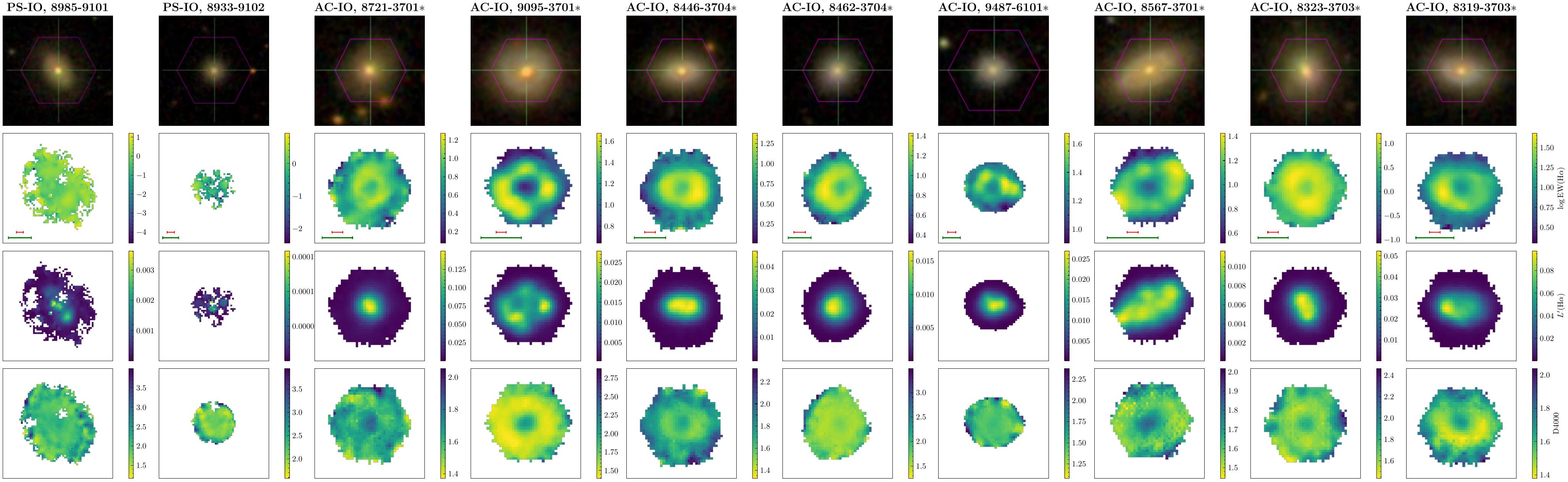}{\textwidth}{}}\vspace*{-2.5\baselineskip}
\gridline{\fig{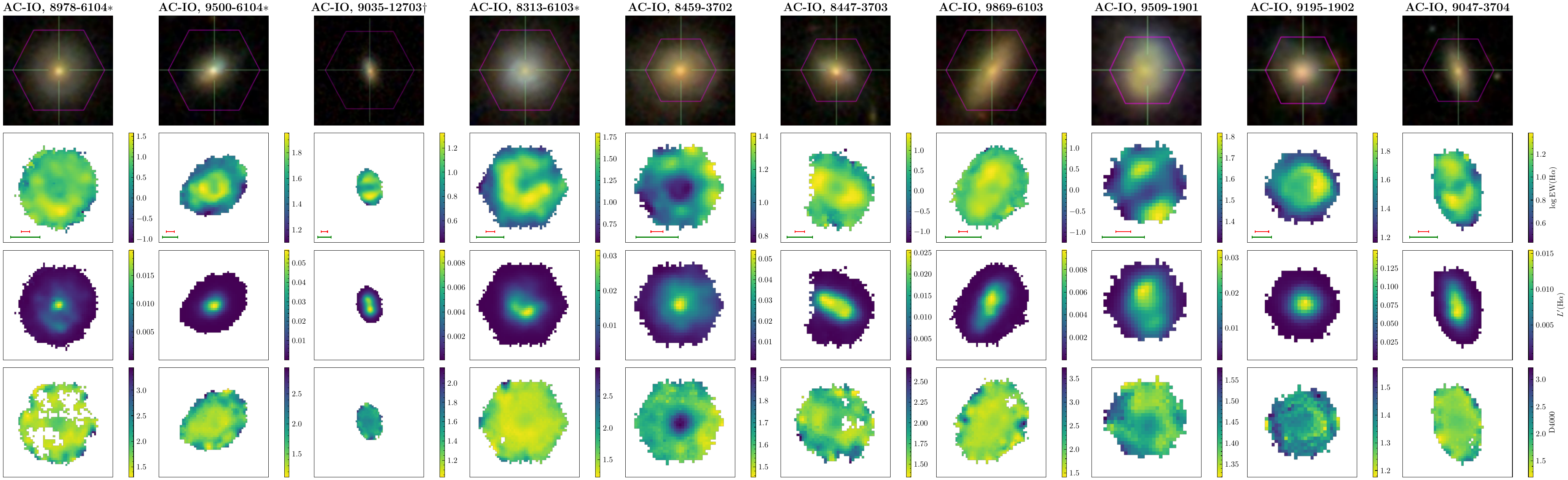}{\textwidth}{}}\vspace*{-2.1\baselineskip}
\caption{Cont.}
\end{figure}

\setcounter{figure}{0}
\begin{figure}[htb!]
\gridline{\fig{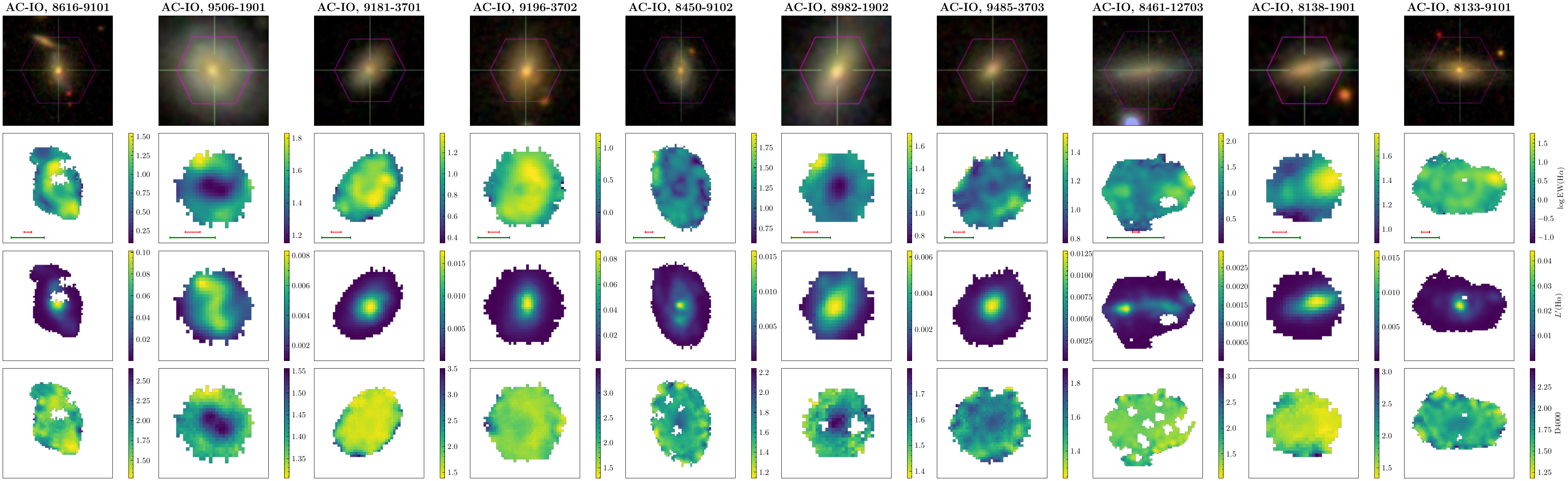}{\textwidth}{}}\vspace*{-2.5\baselineskip}
\gridline{\fig{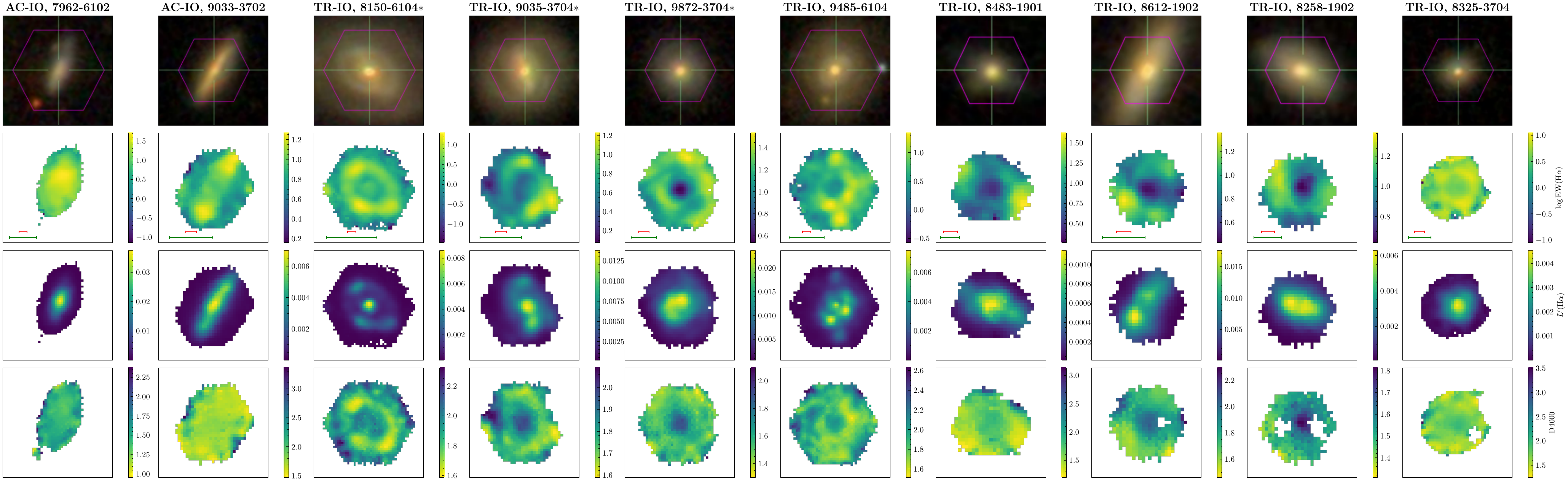}{\textwidth}{}}\vspace*{-2.5\baselineskip}
\gridline{\fig{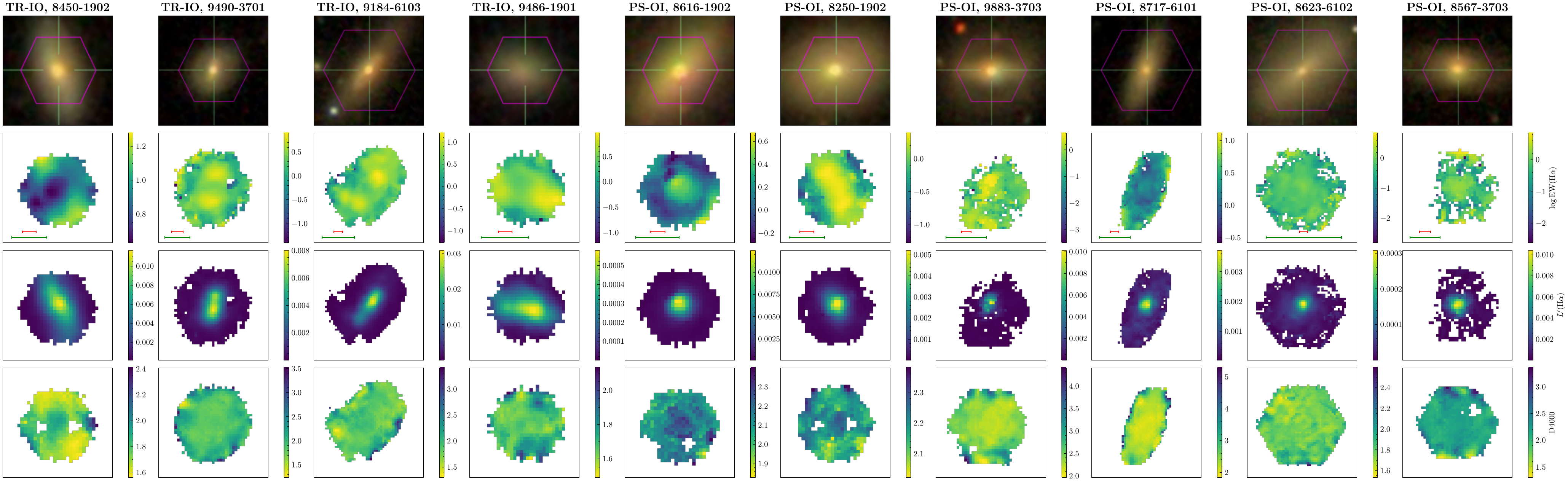}{\textwidth}{}}\vspace*{-2.5\baselineskip}
\gridline{\fig{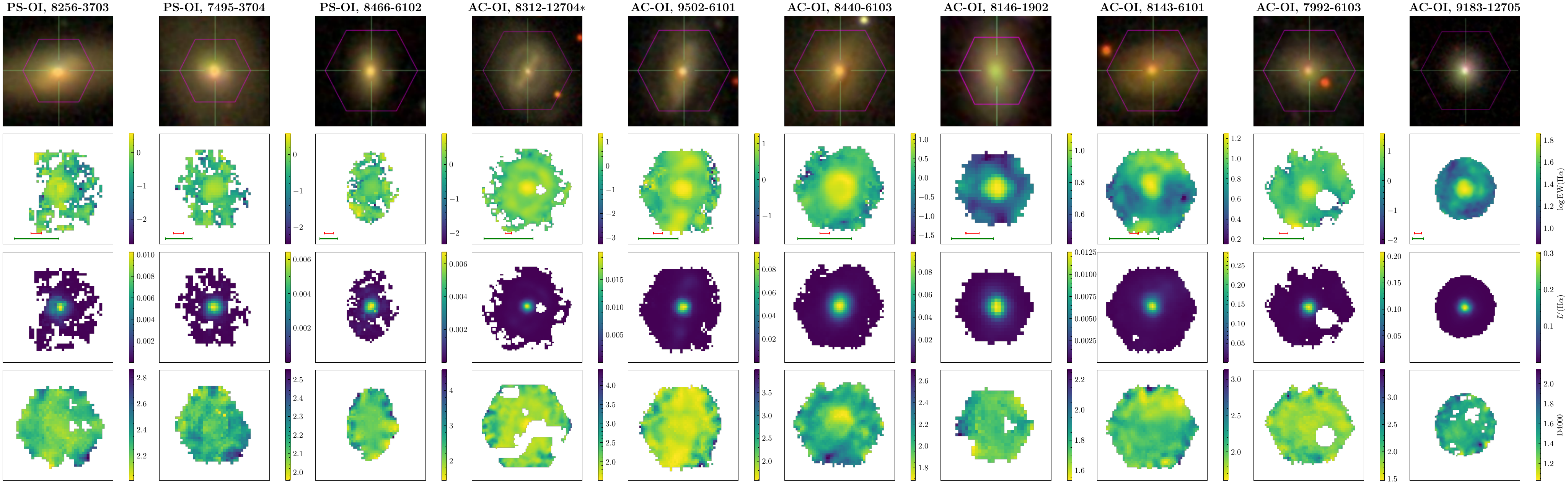}{\textwidth}{}}\vspace*{-2.1\baselineskip}
\caption{Cont.}
\end{figure}

\setcounter{figure}{0}
\begin{figure}[htb!]
\gridline{\fig{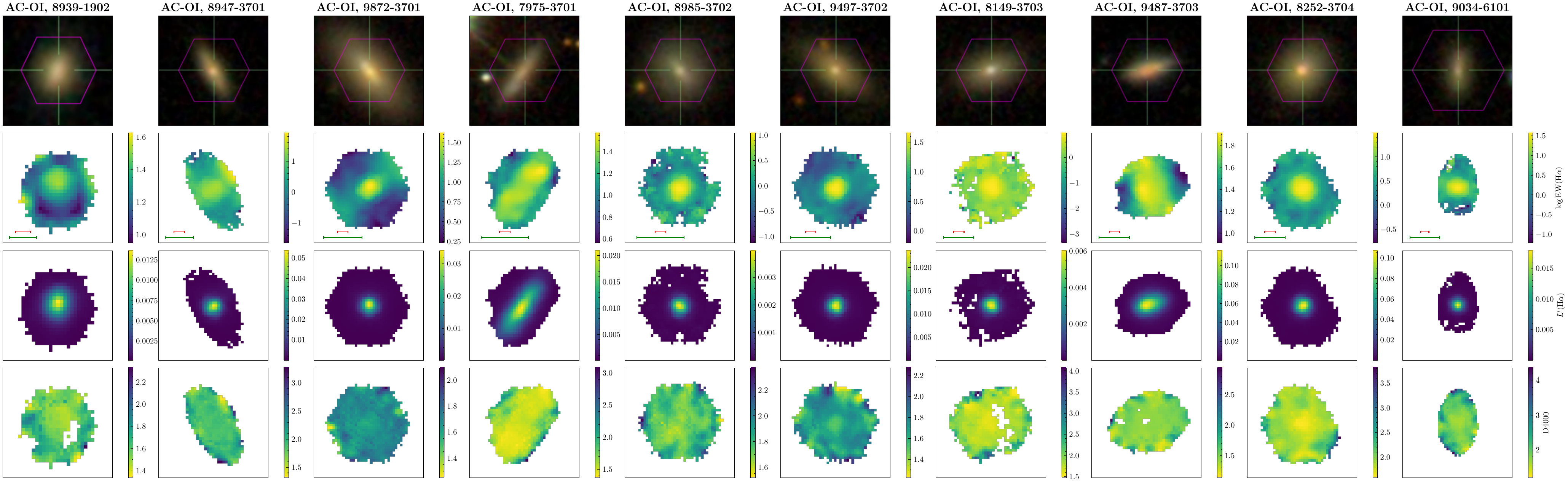}{\textwidth}{}}\vspace*{-2.5\baselineskip}
\gridline{\fig{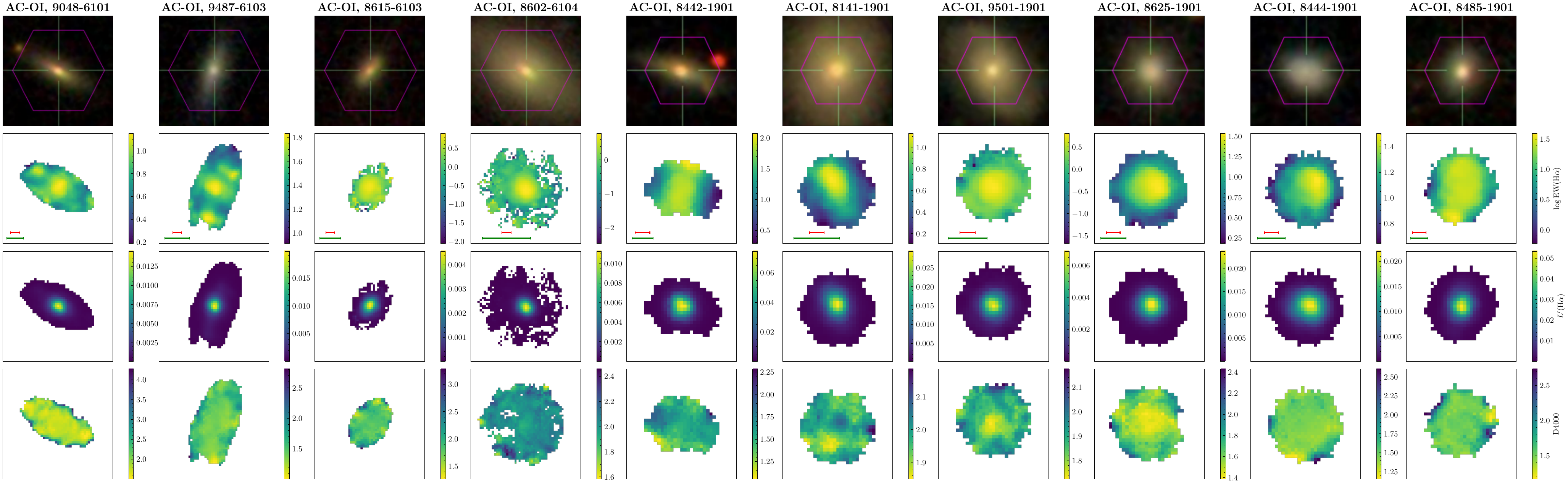}{\textwidth}{}}\vspace*{-2.5\baselineskip}
\gridline{\fig{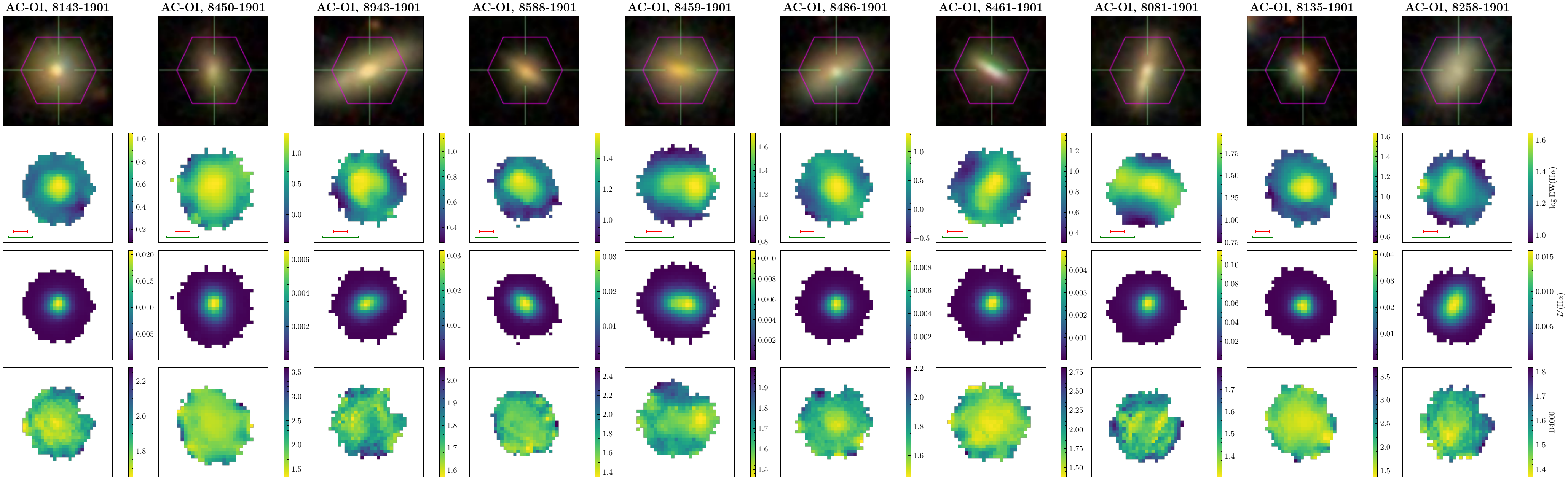}{\textwidth}{}}\vspace*{-2.5\baselineskip}
\gridline{\fig{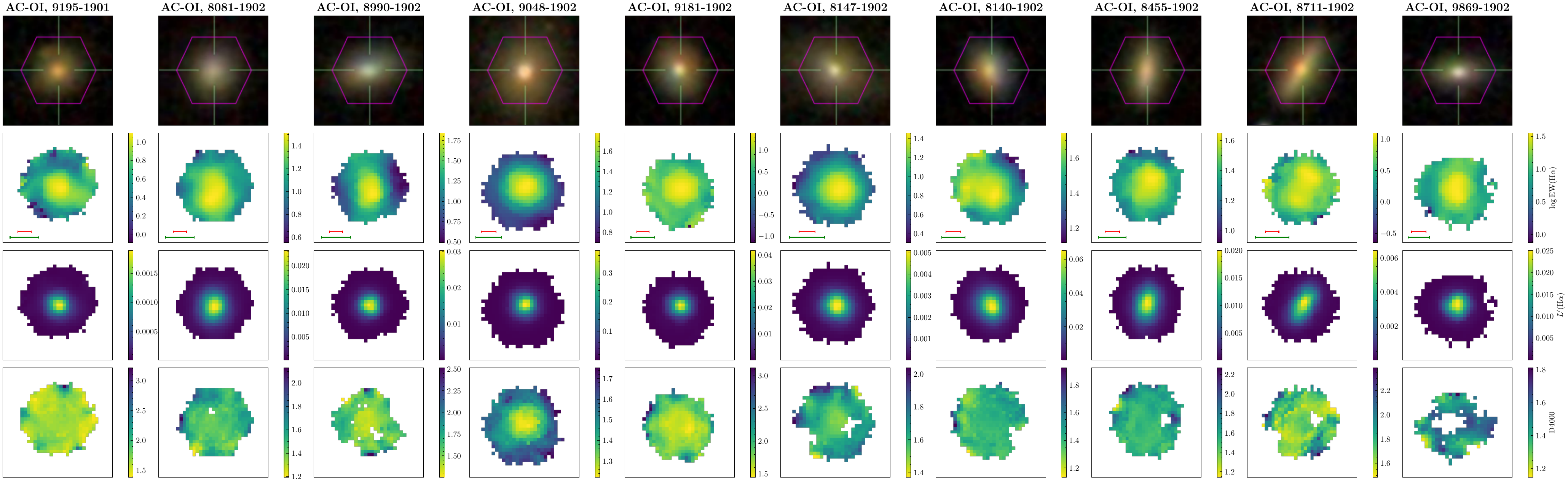}{\textwidth}{}}\vspace*{-2.1\baselineskip}
\caption{Cont.}
\end{figure}

\setcounter{figure}{0}
\begin{figure}[htb!]
\gridline{\fig{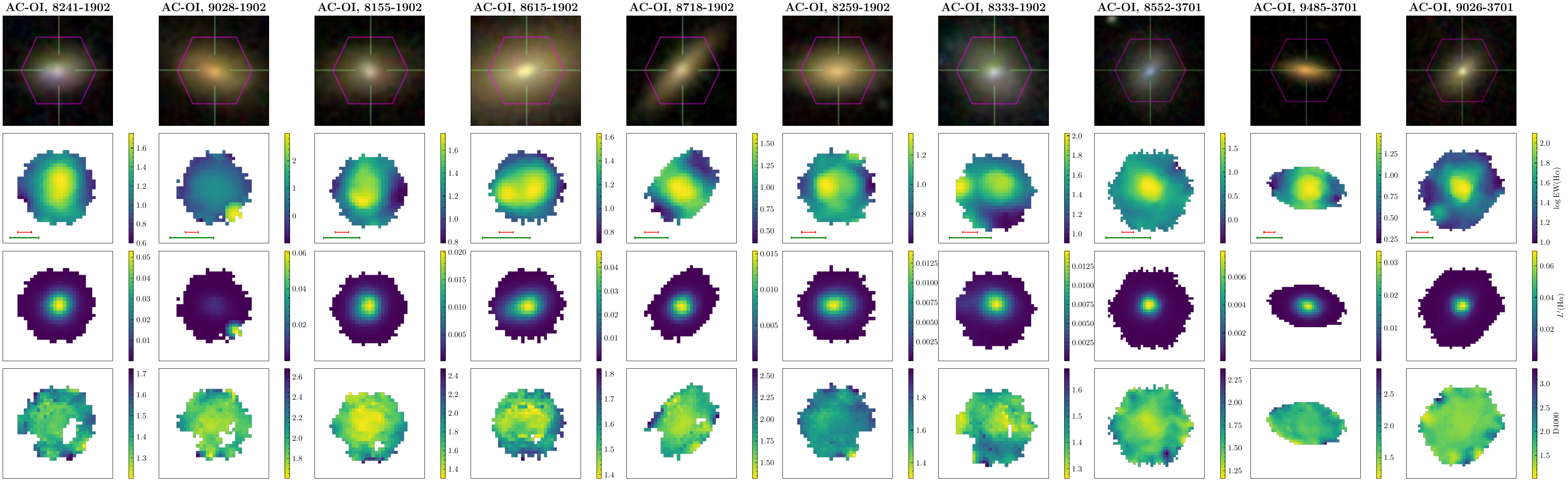}{\textwidth}{}}\vspace*{-2.5\baselineskip}
\gridline{\fig{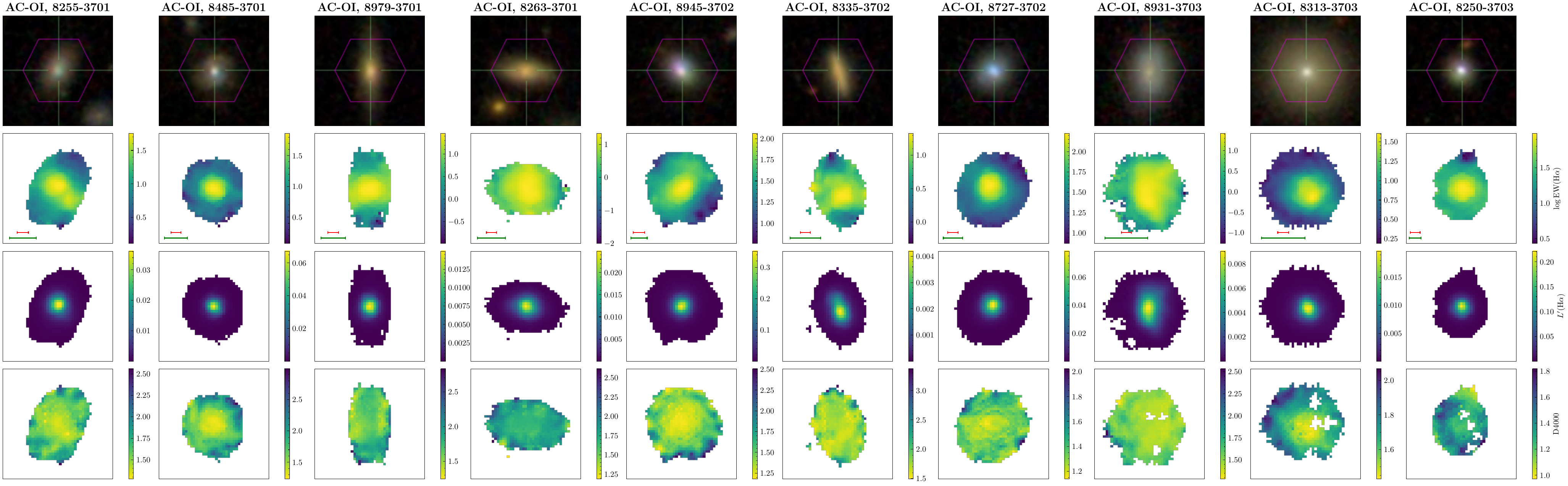}{\textwidth}{}}\vspace*{-2.5\baselineskip}
\gridline{\fig{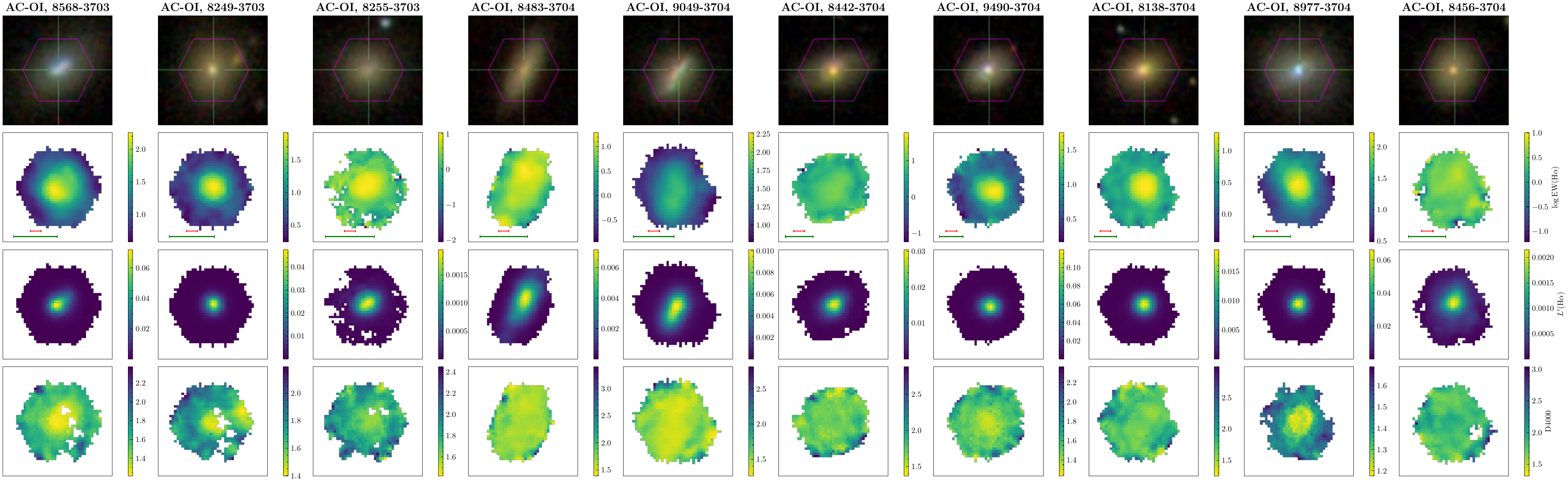}{\textwidth}{}}\vspace*{-2.5\baselineskip}
\gridline{\fig{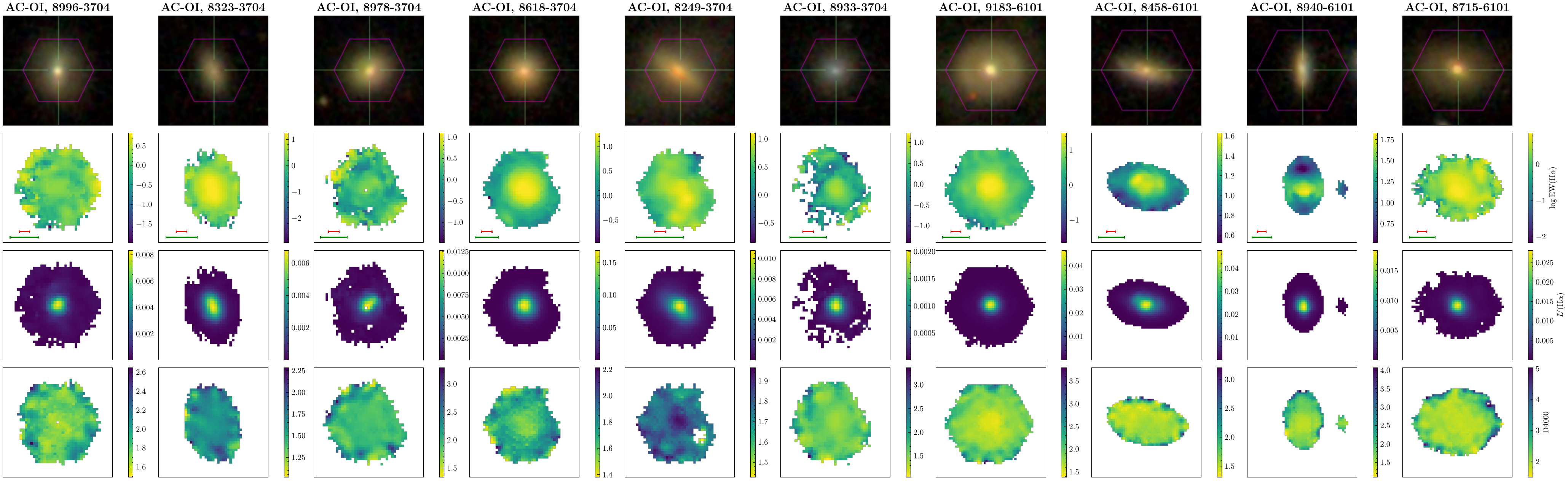}{\textwidth}{}}\vspace*{-2.1\baselineskip}
\caption{Cont.}
\end{figure}

\setcounter{figure}{0}
\begin{figure}[htb!]
\gridline{\fig{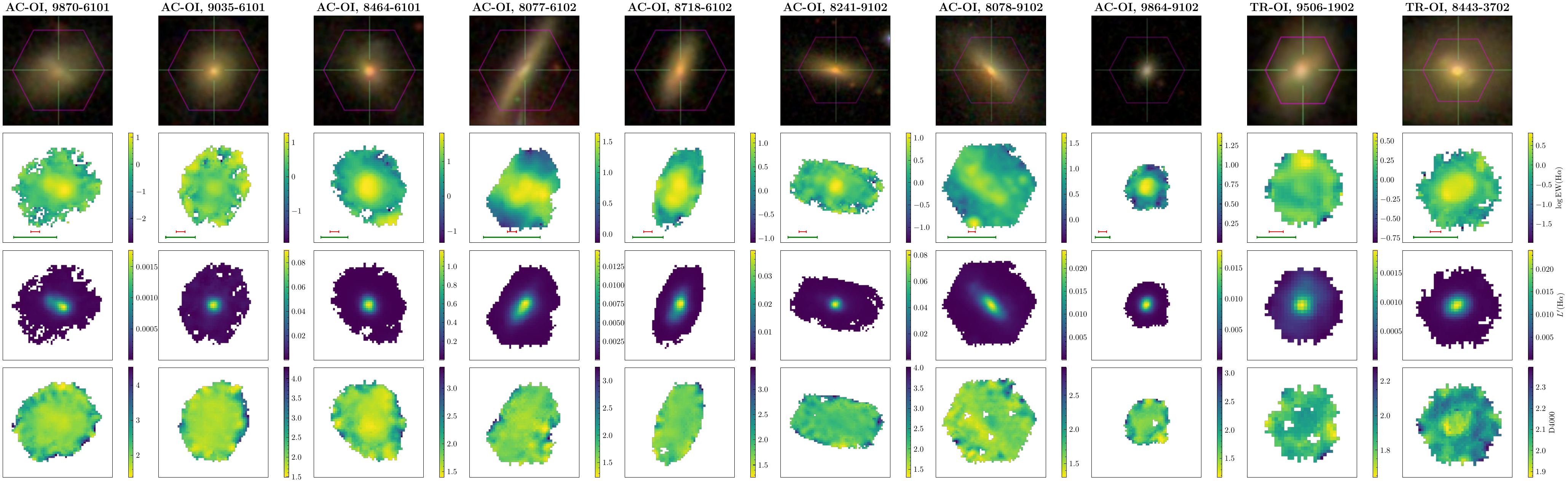}{\textwidth}{}}\vspace*{-2.5\baselineskip}
\gridline{\fig{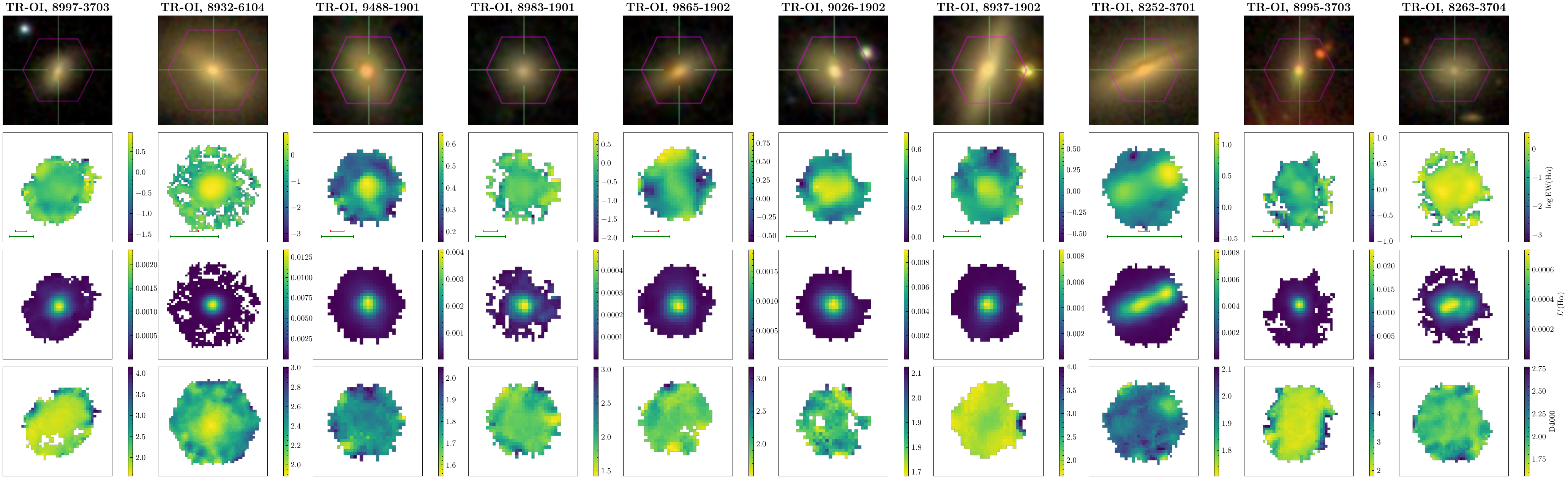}{\textwidth}{}}\vspace*{-2.5\baselineskip}
\gridline{\fig{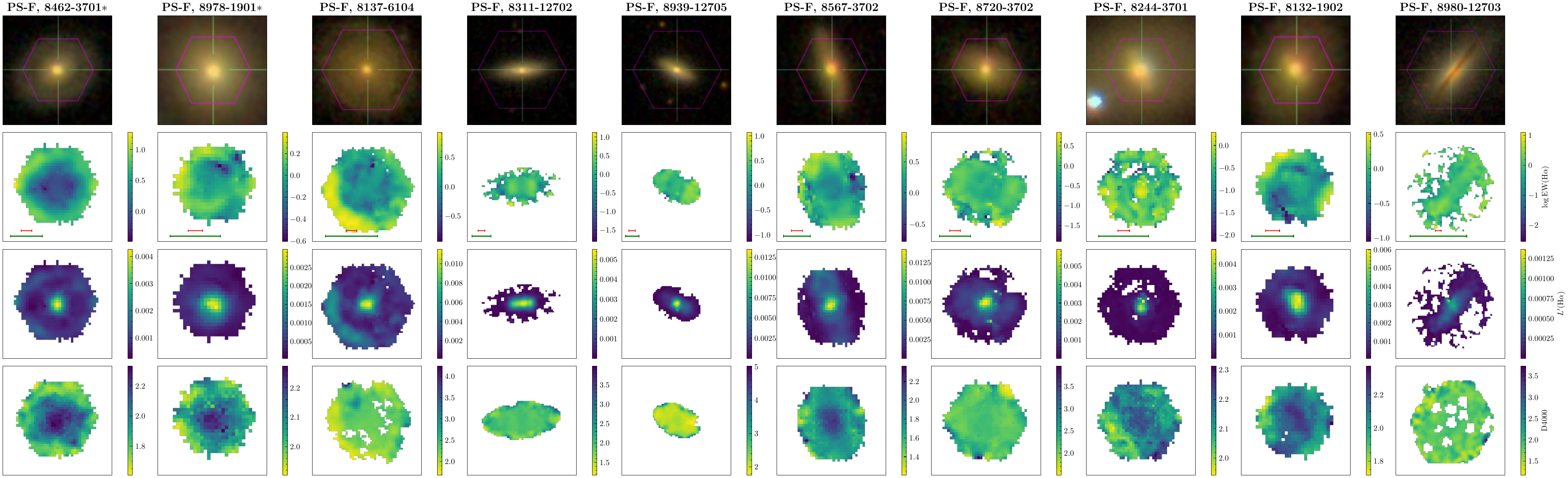}{\textwidth}{}}\vspace*{-2.5\baselineskip}
\gridline{\fig{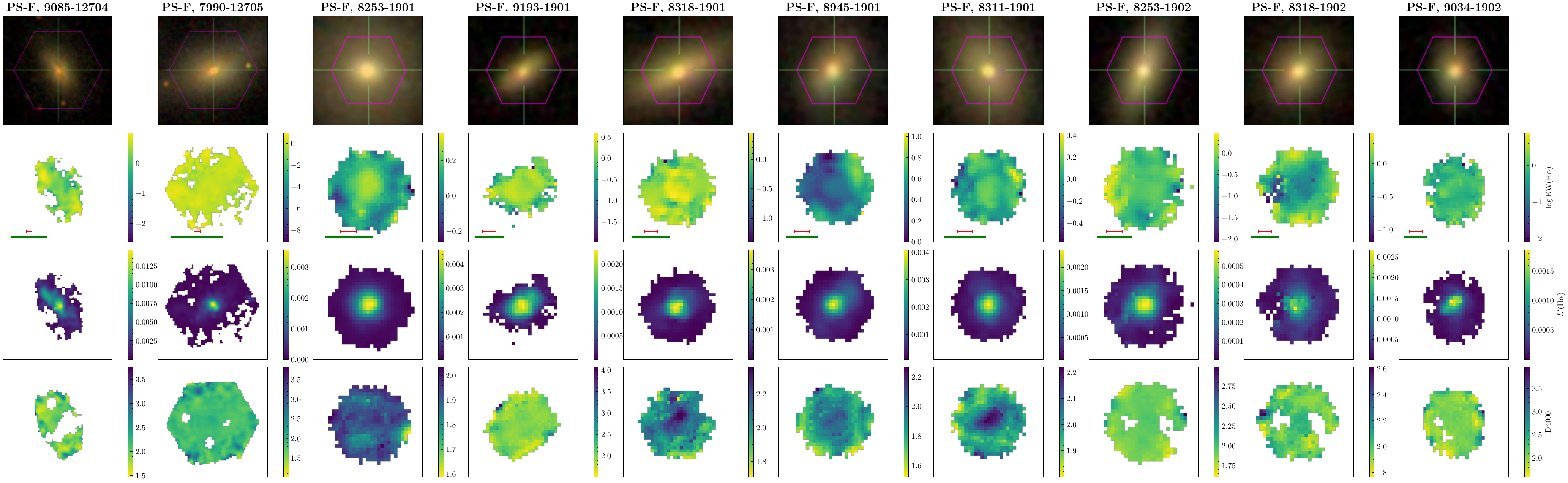}{\textwidth}{}}\vspace*{-2.1\baselineskip}
\caption{Cont.}
\end{figure}

\setcounter{figure}{0}
\begin{figure}[htb!]
\gridline{\fig{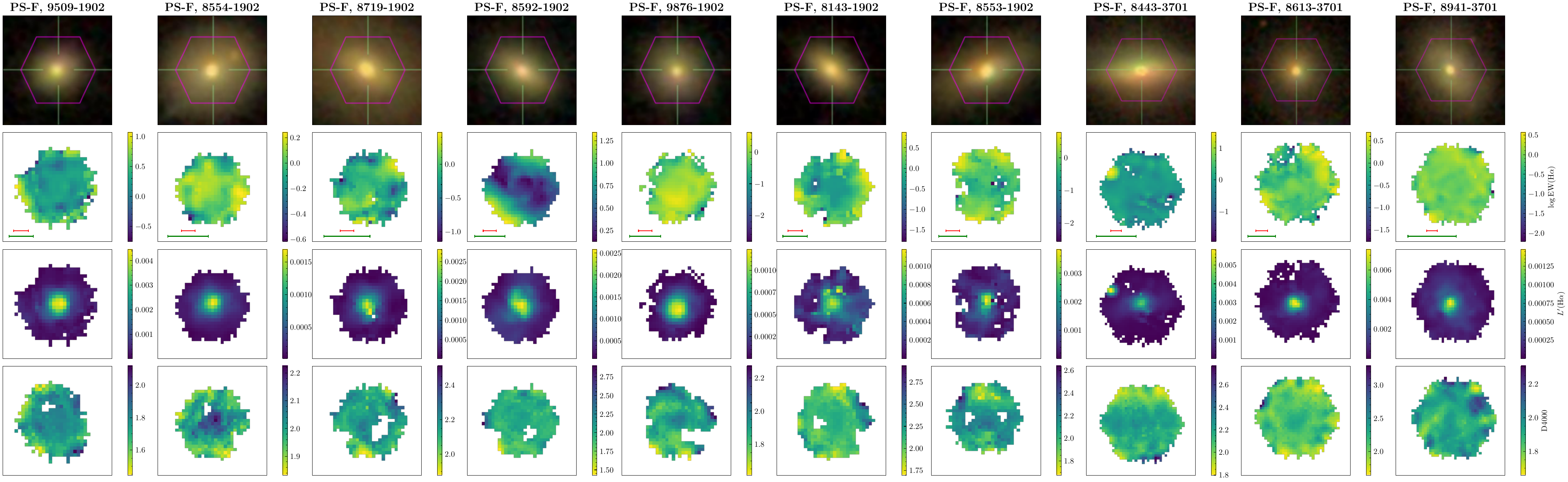}{\textwidth}{}}\vspace*{-2.5\baselineskip}
\gridline{\fig{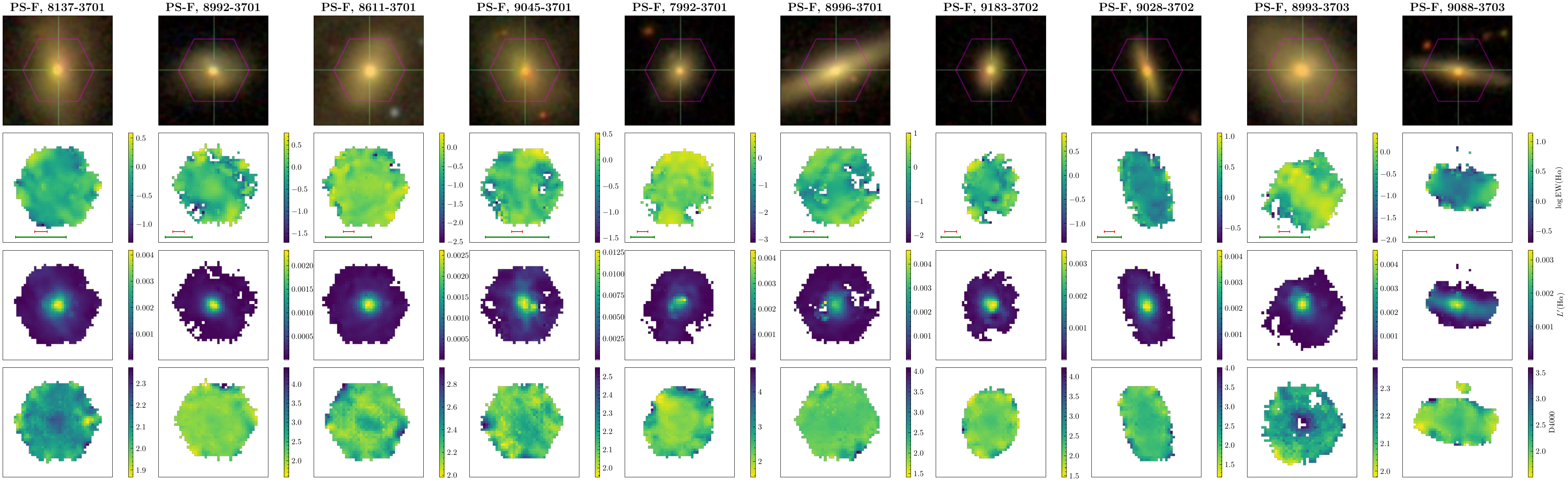}{\textwidth}{}}\vspace*{-2.5\baselineskip}
\gridline{\fig{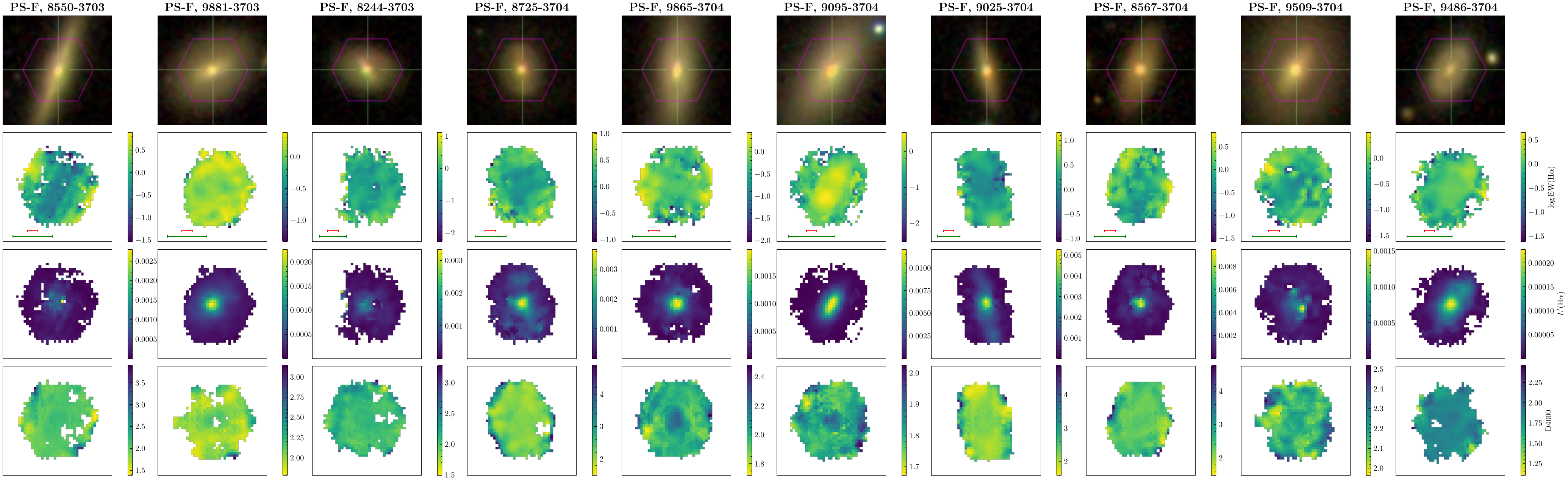}{\textwidth}{}}\vspace*{-2.5\baselineskip}
\gridline{\fig{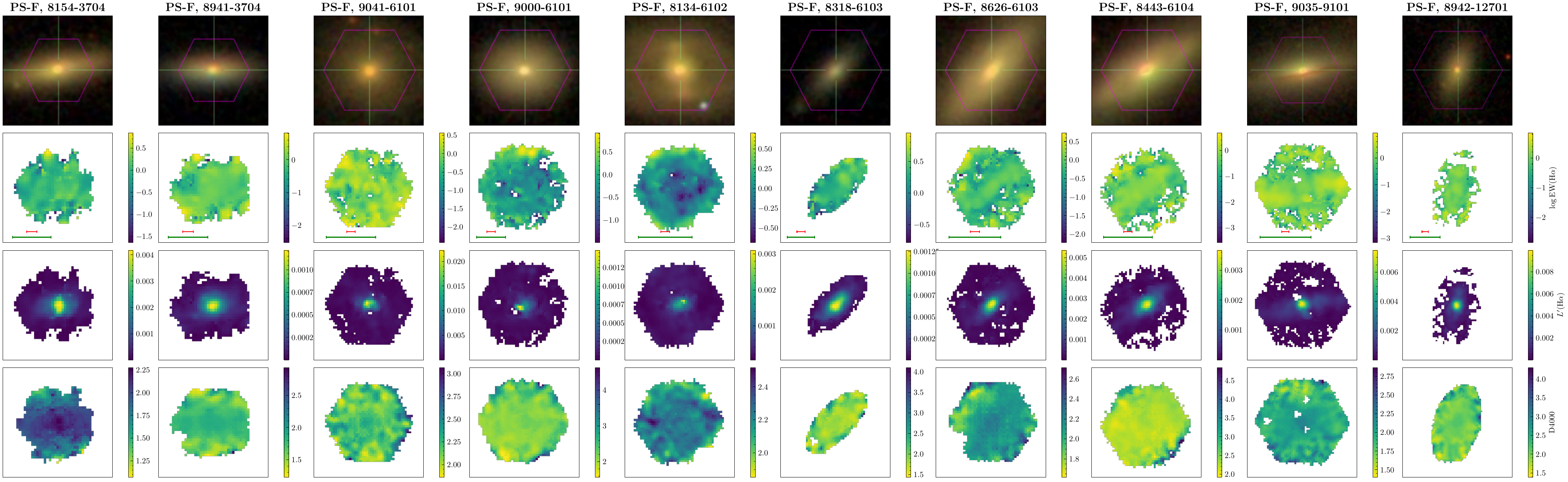}{\textwidth}{}}\vspace*{-2.1\baselineskip}
\caption{Cont.}
\end{figure}

\setcounter{figure}{0}
\begin{figure}[htb!]
\gridline{\fig{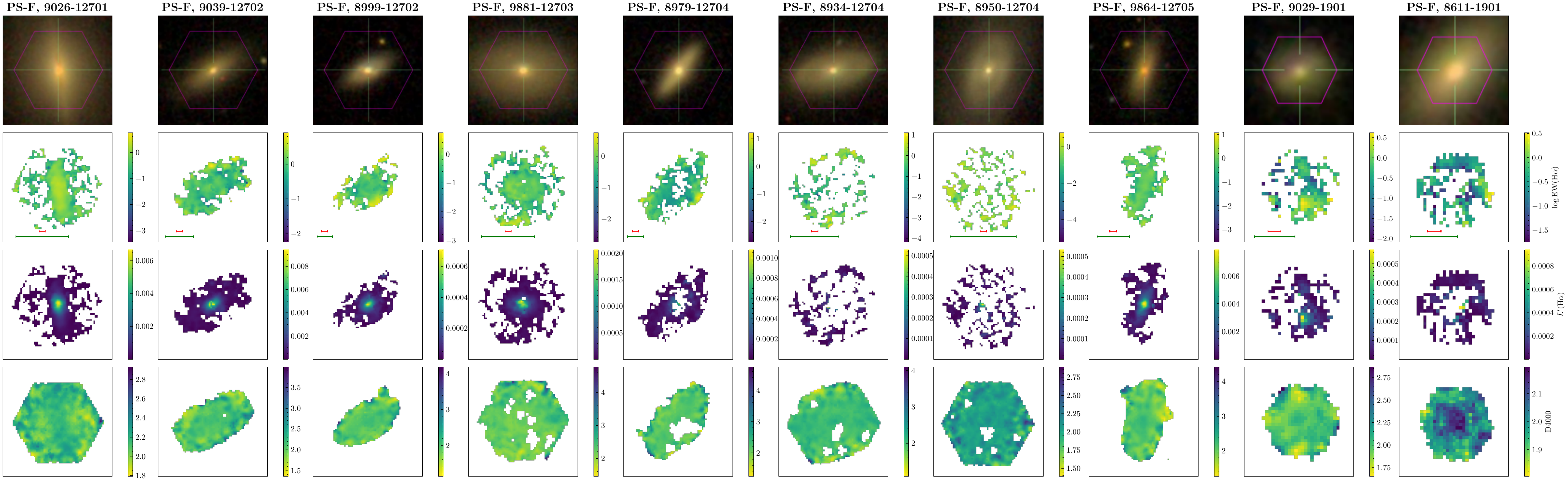}{\textwidth}{}}\vspace*{-2.5\baselineskip}
\gridline{\fig{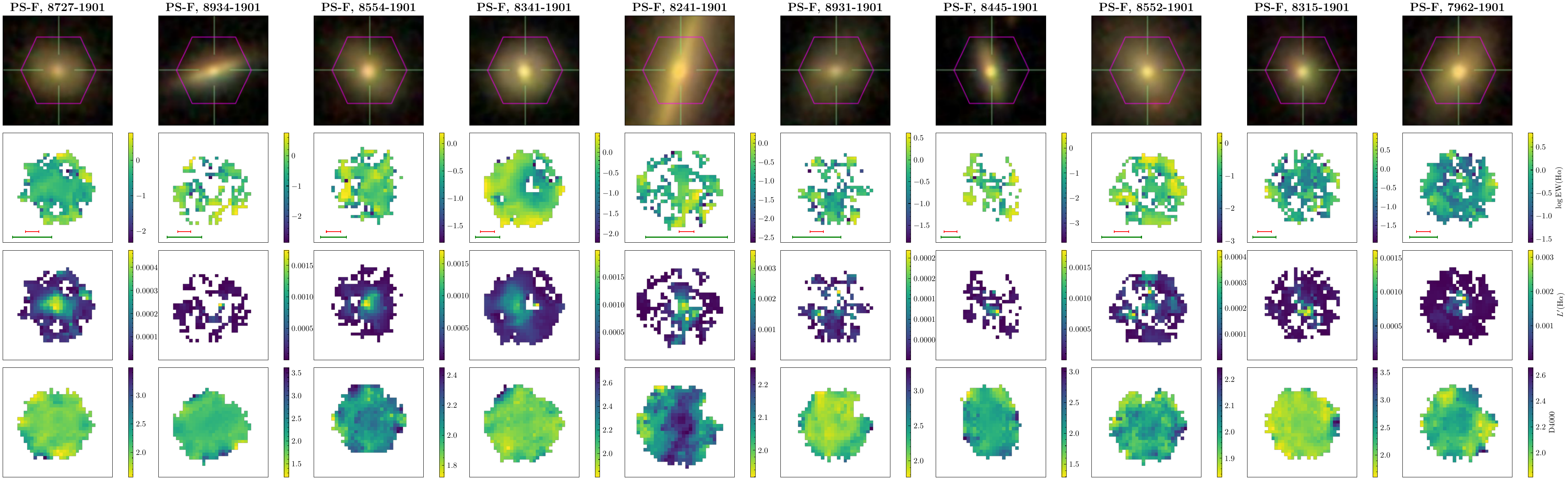}{\textwidth}{}}\vspace*{-2.5\baselineskip}
\gridline{\fig{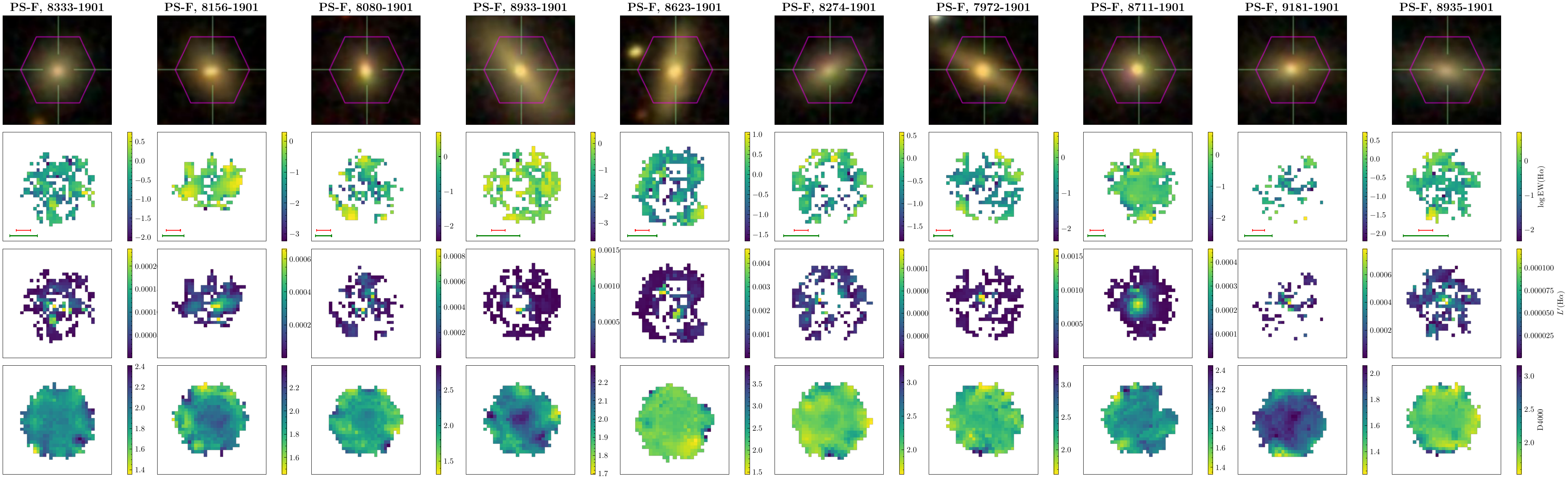}{\textwidth}{}}\vspace*{-2.5\baselineskip}
\gridline{\fig{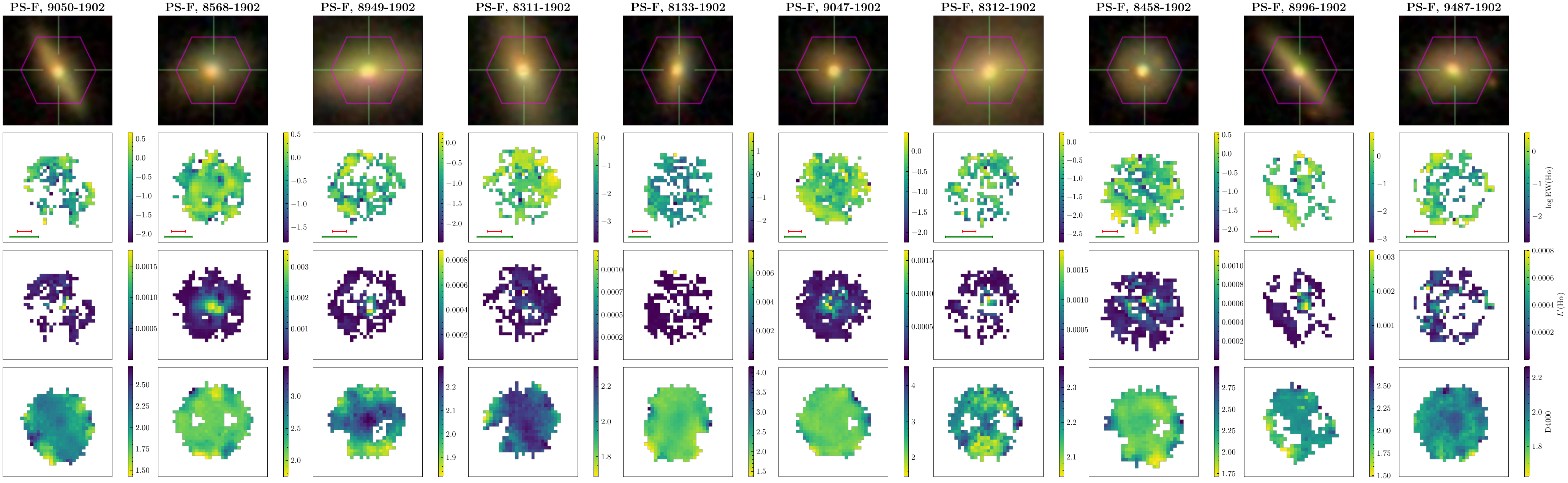}{\textwidth}{}}\vspace*{-2.1\baselineskip}
\caption{Cont.}
\end{figure}

\setcounter{figure}{0}
\begin{figure}[htb!]
\gridline{\fig{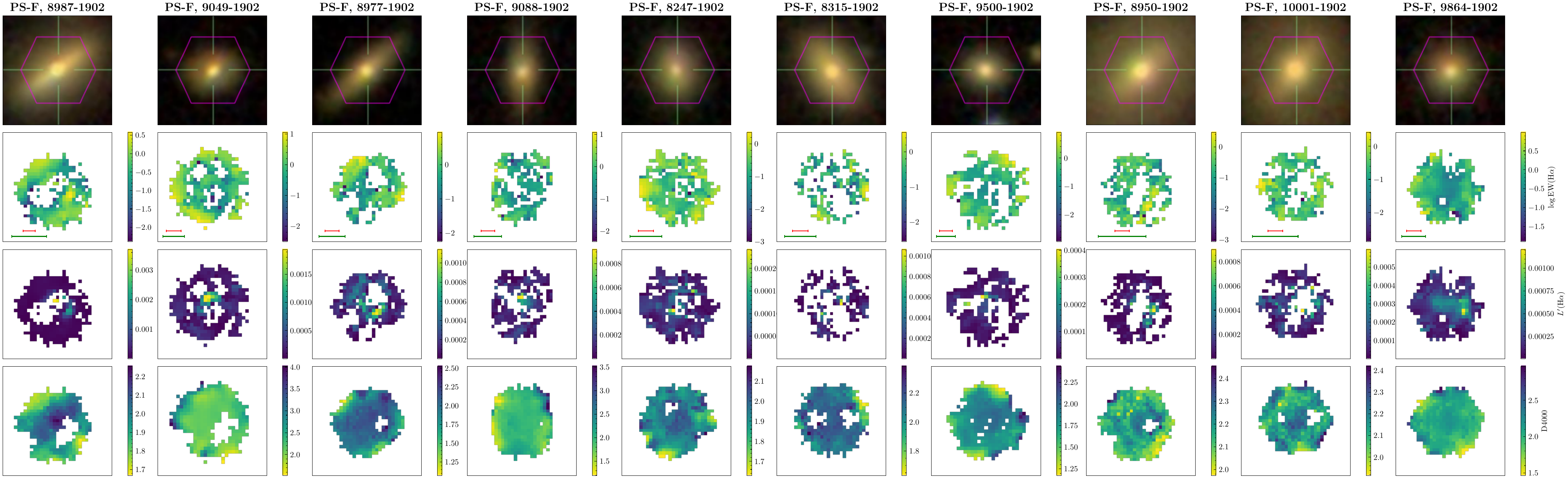}{\textwidth}{}}\vspace*{-2.5\baselineskip}
\gridline{\fig{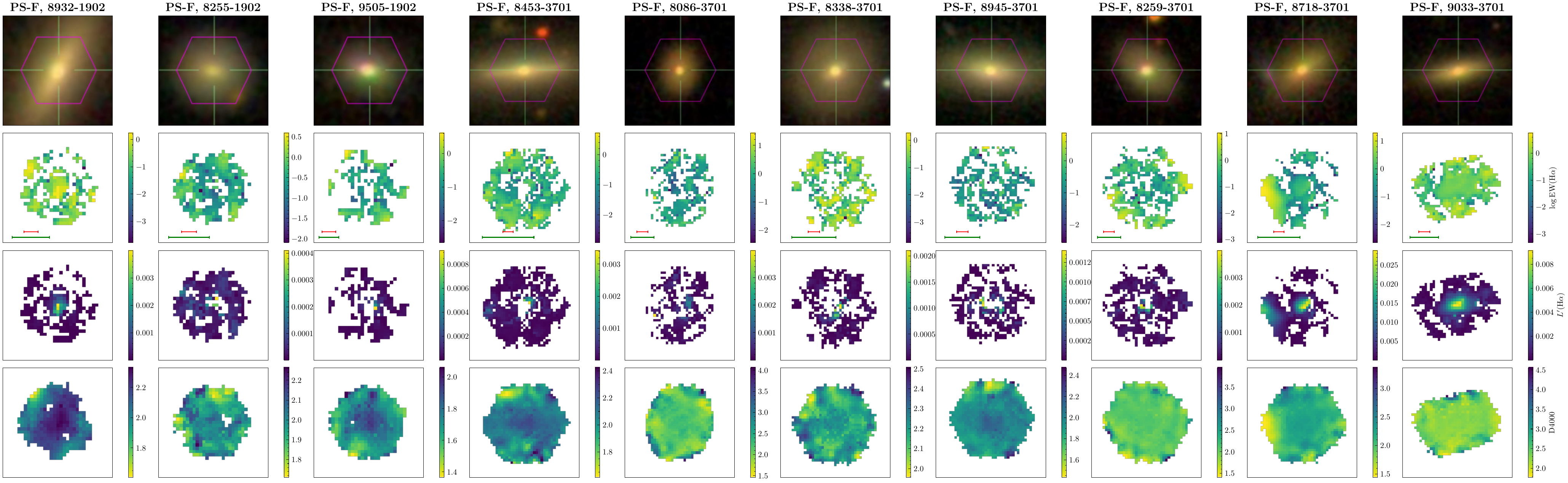}{\textwidth}{}}\vspace*{-2.5\baselineskip}
\gridline{\fig{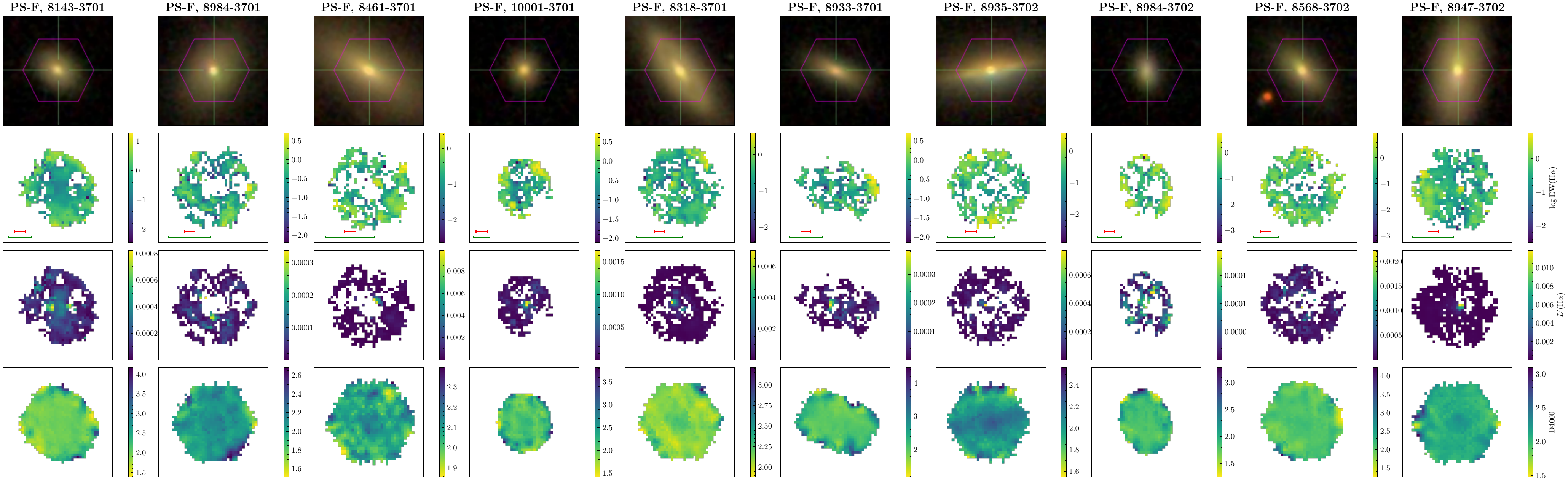}{\textwidth}{}}\vspace*{-2.5\baselineskip}
\gridline{\fig{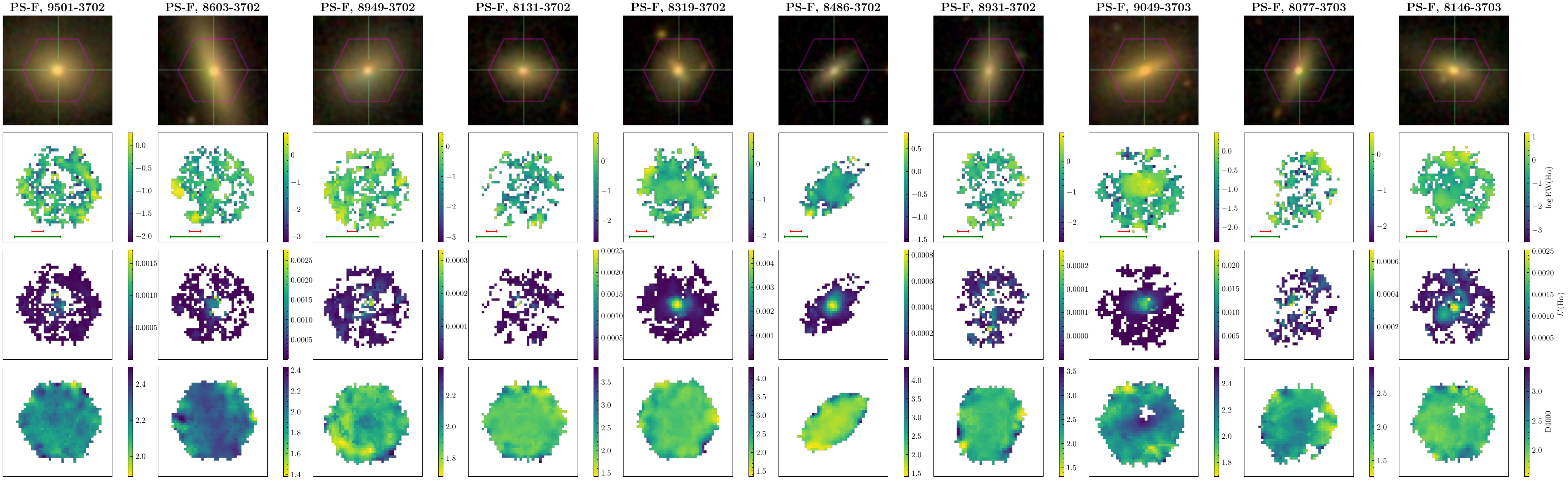}{\textwidth}{}}\vspace*{-2.1\baselineskip}
\caption{Cont.}
\end{figure}

\setcounter{figure}{0}
\begin{figure}[htb!]
\gridline{\fig{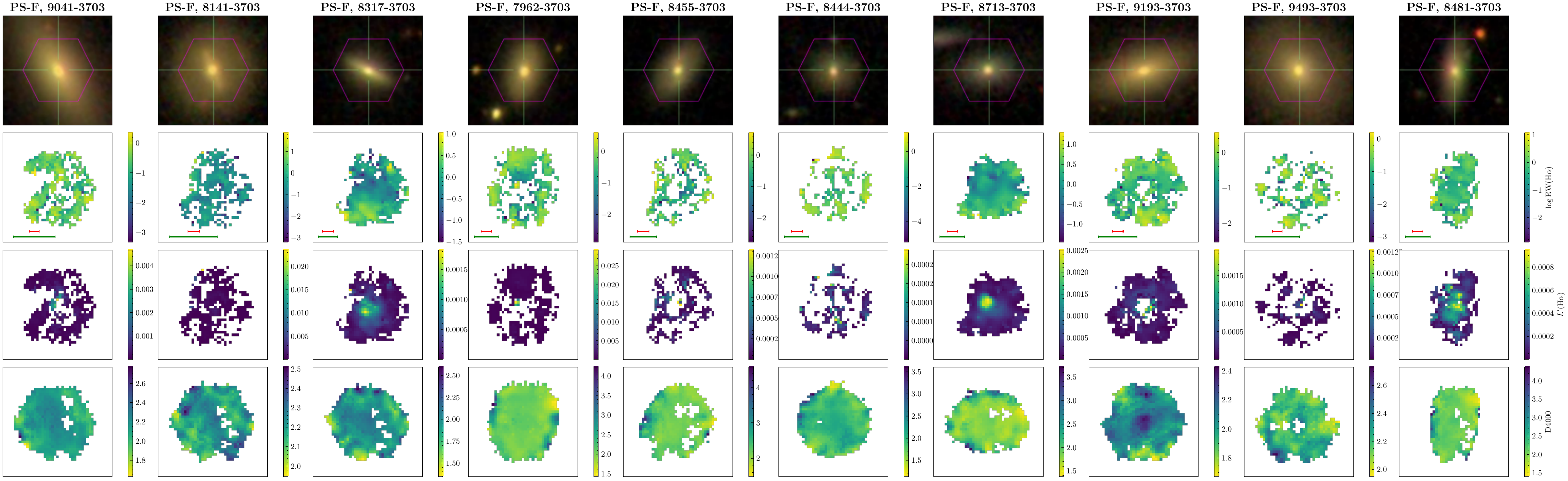}{\textwidth}{}}\vspace*{-2.5\baselineskip}
\gridline{\fig{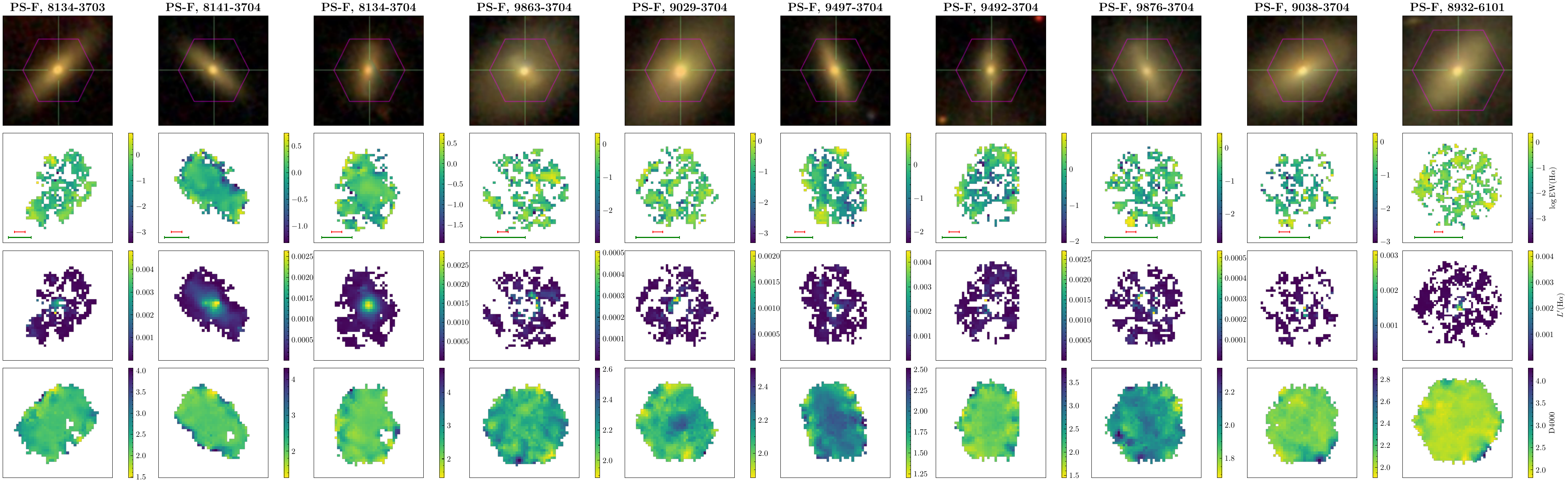}{\textwidth}{}}\vspace*{-2.5\baselineskip}
\gridline{\fig{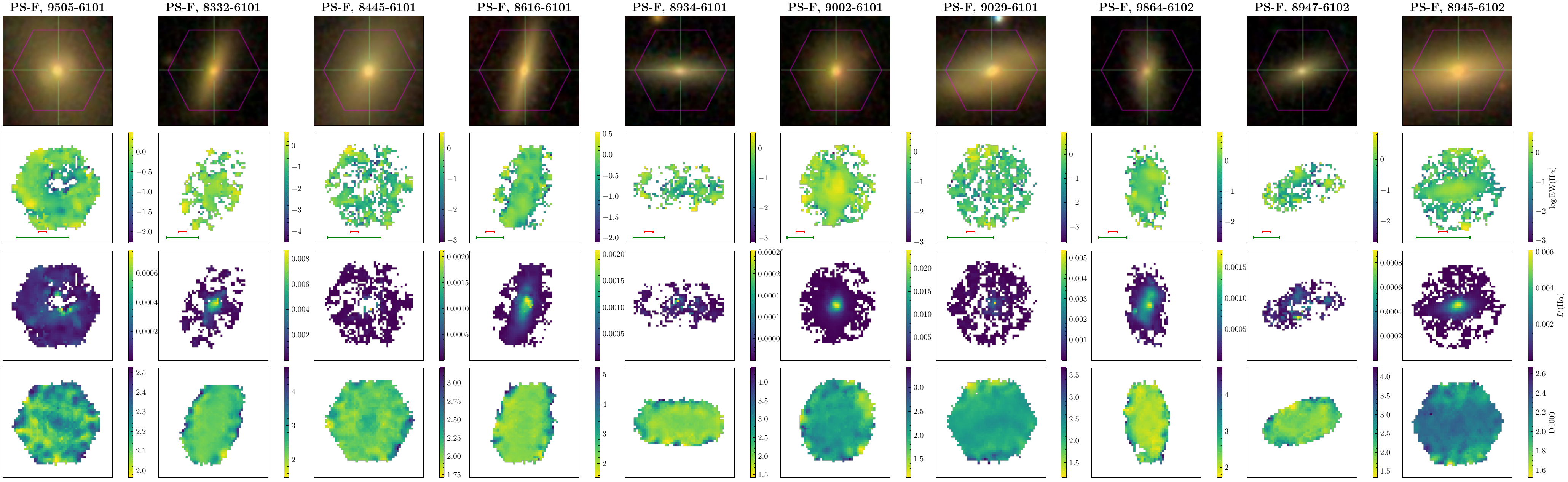}{\textwidth}{}}\vspace*{-2.5\baselineskip}
\gridline{\fig{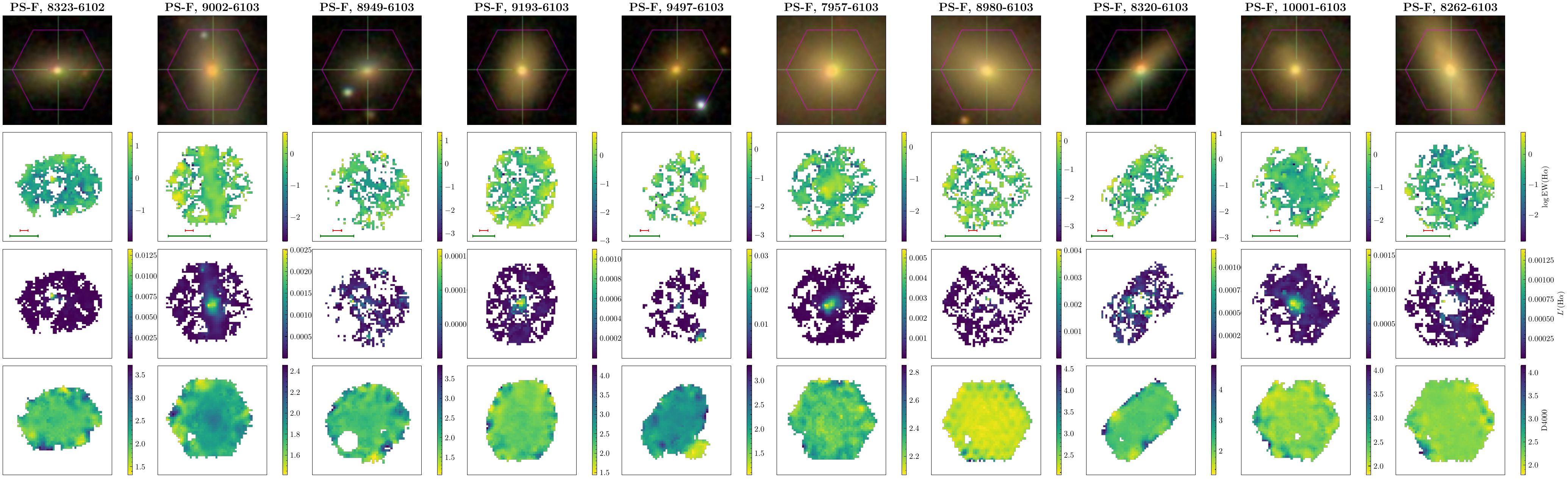}{\textwidth}{}}\vspace*{-2.1\baselineskip}
\caption{Cont.}
\end{figure}

\setcounter{figure}{0}
\begin{figure}[htb!]
\gridline{\fig{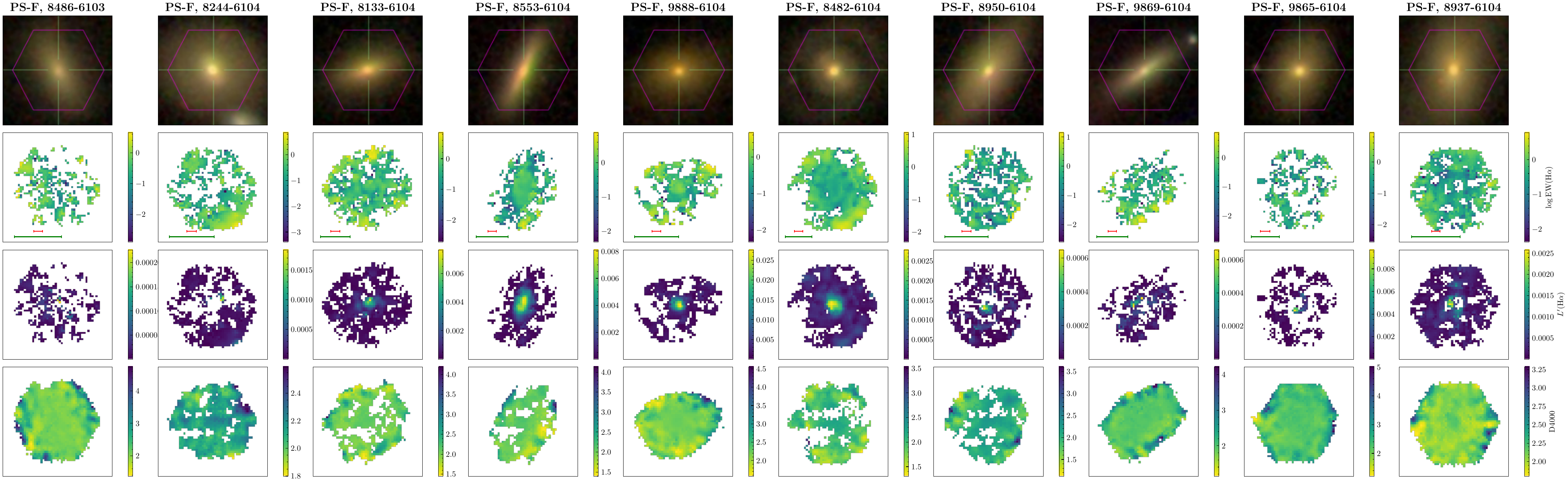}{\textwidth}{}}\vspace*{-2.5\baselineskip}
\gridline{\fig{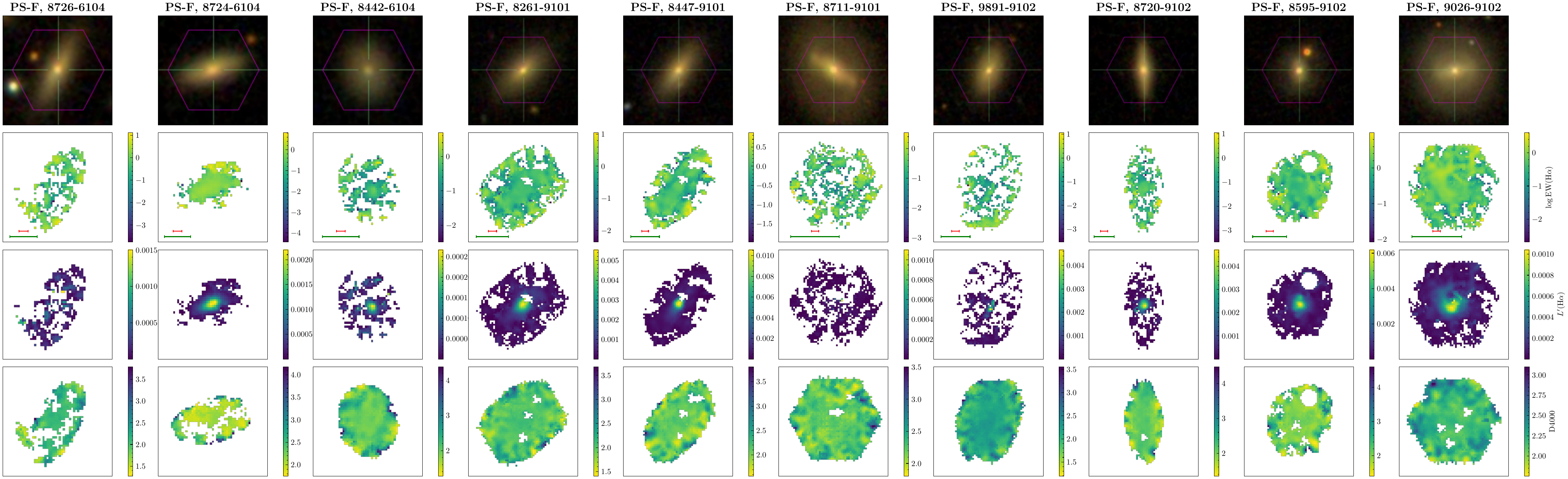}{\textwidth}{}}\vspace*{-2.5\baselineskip}
\gridline{\fig{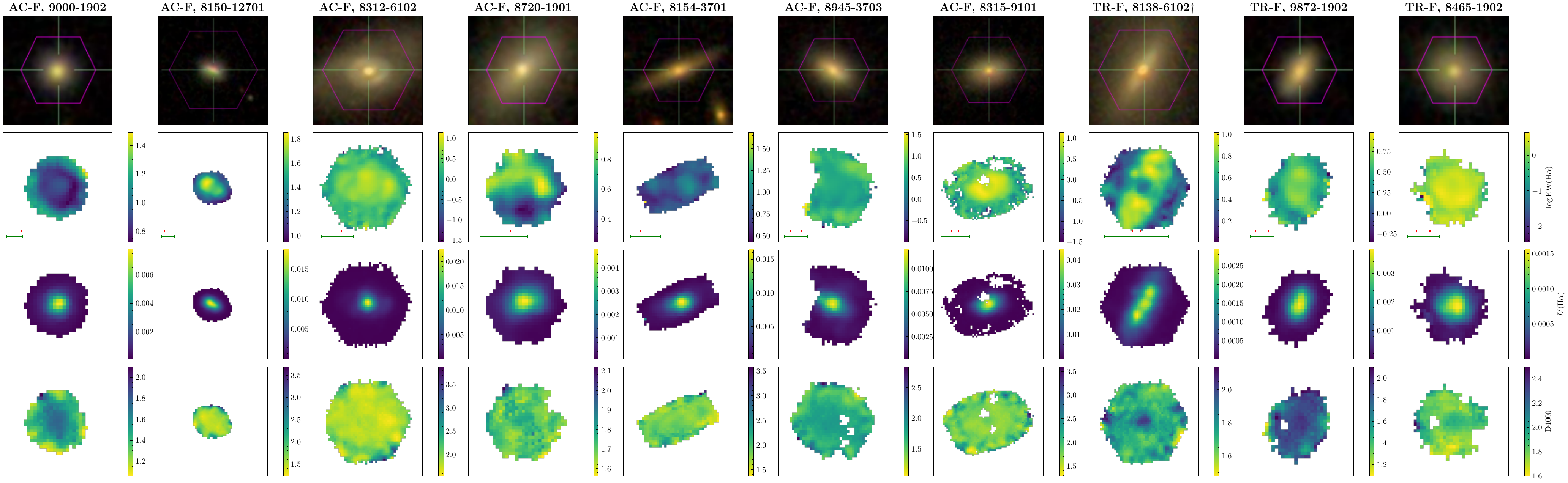}{\textwidth}{}}\vspace*{-2.5\baselineskip}
\gridline{\fig{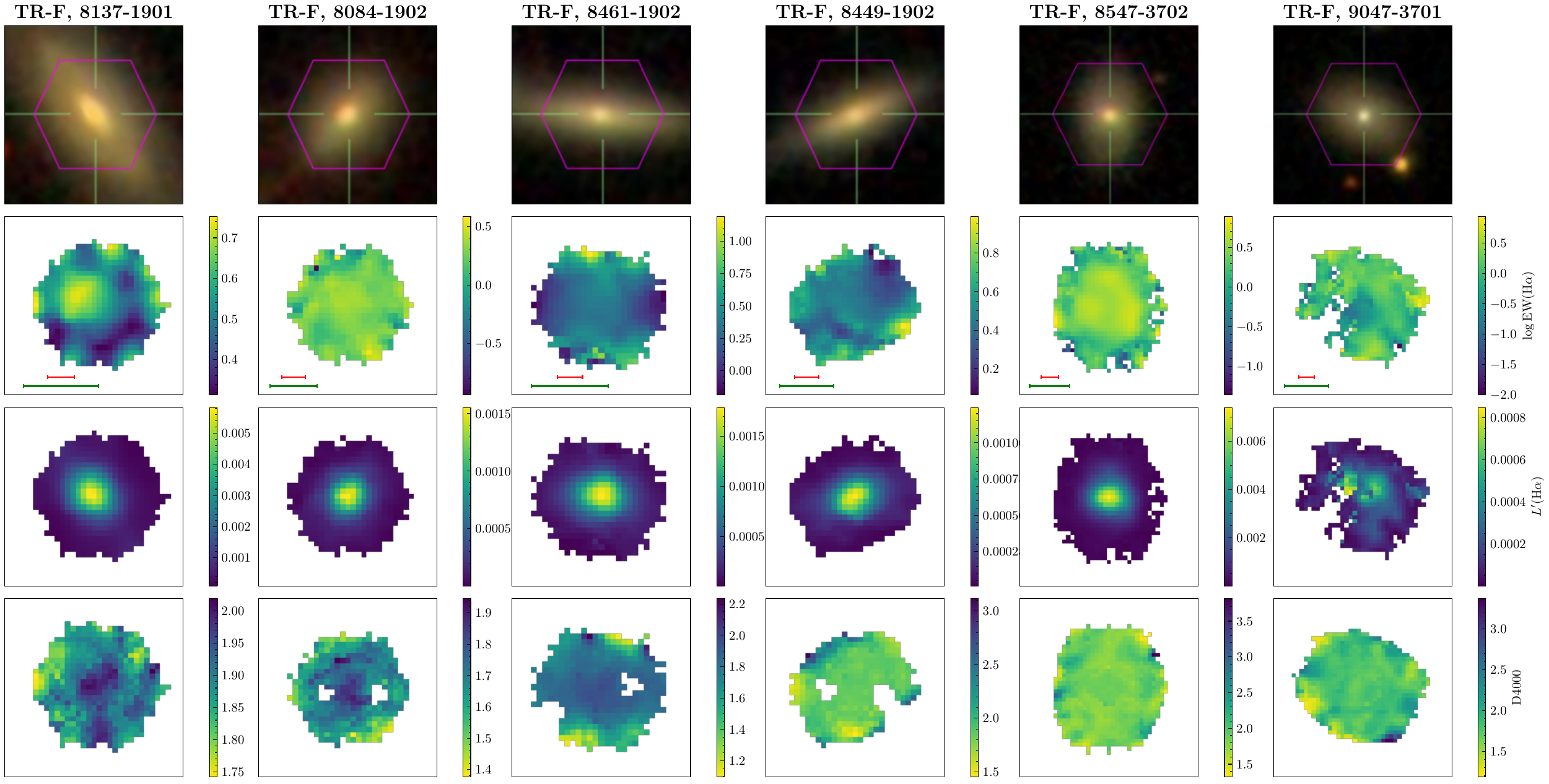}{0.6\textwidth}{}}\vspace*{-2.1\baselineskip}
\caption{Cont.}
\end{figure}

\end{document}